\documentclass[aps,prd,amssymb,groupedaddress,nofootinbib,notitlepage]{revtex4-1}

\usepackage{epsfig}
\usepackage[english]{babel}
\usepackage{amsfonts}
\usepackage{amssymb}
\usepackage{amsmath}
\usepackage{multirow}
\usepackage[latin1]{inputenc}
\usepackage{verbatim}
\usepackage{type1cm}
\usepackage{setspace}
\usepackage[FIGTOPCAP]{subfigure}
\def\ind{\mathfrak{l}}
\def\vec{\boldsymbol}
\def\mat{\boldsymbol}

\def\setsize{\csname @setfontsize\endcsname \setsize}

\begin{document}


\title{Towards  efficient and  optimal analysis of CMB anisotropies on a masked sky}

\author{H.\,F. Gruetjen$^{1}$}
\author{E.\,P.\,S. Shellard$^{1}$}

\affiliation
{$^{1}$Centre for Theoretical Cosmology,
DAMTP,
University of Cambridge,
 CB3 0WA, UK}

\date{\today}

\begin{abstract}
Strong foreground contamination in high resolution CMB data requires masking which introduces statistical anisotropies and renders a full maximum likelihood analysis numerically intractable. Standard analysis methods like the pseudo-$C_l$ framework lead to information loss due to estimator suboptimalities. We set out and validate a methodology for numerically efficient estimators for a masked sky that recover nearly as much information as a full maximum likelihood procedure. In addition to the standard pseudo-$C_l$ statistic, the approach introduces an augmented basis designed to account for the mode coupling due to the masking of the sky. We motivate the choice of this basis by describing the basic structure of the covariance matrix. We demonstrate that the augmented estimator can achieve near-optimal results in the presence of a WMAP-realistic mask.\\
\end{abstract}

\maketitle

\twocolumngrid

\tableofcontents

\section{\label{sec:Intro}Introduction}
The success of high resolution CMB experiments like WMAP has opened up a wealth of new possibilities to test cosmological models of structure formation. The large amount of data has been used to place strong constraints on parameters of statistical models for the CMB temperature fluctuations. Developing accurate and numerically feasible estimators that can optimally extract cosmological information from the data is a central problem in CMB analysis.

Let us denote the deviation from the mean CMB temperature observed in direction $\hat{n}$ by $\Delta T(\hat{n})$. To a very good approximation $\Delta T(\hat{n})$ defines a Gaussian random field on the sphere with zero mean. We will assume for now that we can neglect any potential sources of non-Gaussianities, so the random field is entirely determined by its 2-point function. It is convenient to treat the problem in multipole space and define multipole coefficients of the temperature fluctuations via\footnote{The superscript $R$ in $Y^R_{lm}$ indicates that we are using real spherical harmonics throughout this work to keep the multipole coefficients of the temperature fluctuations manifestly real.}
\begin{equation}
a_{lm}=\int\mathrm{d}\hat{n}\,Y^R_{lm}(\hat{n})\Delta T(\hat{n})
\end{equation}
Assuming the universe to be statistically isotropic, we also expect the random field $\Delta T(\hat{n})$ to be statistically isotropic. It can be shown that this implies
\begin{equation}
\langle a_{l_1m_1}a_{l_2m_2}\rangle=C_{l_1}\delta_{l_1l_2}\delta_{m_1m_2}
\end{equation}
where $C_{l}$ is the power spectrum of temperature fluctuations. Hence the statistics of the temperature fluctuations are entirely determined by their power spectrum and cosmological parameters influence the fluctuations solely through the power spectrum. 

A common technique to extract information from the CMB is to estimate parameters of the statistical model using Maximum Likelihood (ML) estimation which will be discussed in more detail below (see for example \cite{TegmarkBunn:bruteforce, Gorski:COBEanalysis, Borrill:MADCAP, BondJaffeKnox:estCMB, Tegmark:losslessmeasure} for applications of ML estimation to CMB analysis). ML estimation for Gaussian distributions generically requires in some way the inversion of the covariance matrix and a weighting of the data with this inverse. The discussion above suggests that this is not a problem as the covariance structure of the multipole coefficients seems to be extremely simple. It is merely a diagonal matrix that can be trivially inverted. Unfortunately the true multipole coefficients in the above discussion are not accessible in practice and the covariance structure of experimental multipole coefficients is much more complicated. This is due to two reasons. First, anisotropic instrumental noise contributes to the measured multipole coefficient and leads to correlations between them. Secondly, foreground contamination due to radiation sources both within our on galaxy and beyond make it necessary to mask the sky with certain templates. The spherical harmonics are not orthonormal on an incomplete sky and hence the introduction of the mask leads to significant mode coupling and cross correlations between the multipole coefficients.

These two effects have to be taken into account in the analysis and they turn the covariance matrix into a dense complicated matrix. Unfortunately, this means that inversion becomes intractable for high resolution experiments. These analyse the CMB up to multipoles with $l_{\mathrm{max}}$ of a few times $10^3$ implying that for exact ML methods the inversion of a huge $10^7\times 10^7$ matrix is required which is numerically not feasible at least by direct calculation.

Conjugate gradient methods \cite{Smith:GravLensInvVarweight} or iterative procedures \cite{Elsner:EffWiener} can be employed to achieve inverse covariance weighting of the data under certain albeit not too restrictive conditions and can potentially lead to feasible routes to an exact calculation of the Fisher matrix \cite{Elsner:FastFishCalc, Elsner:LikeFishSys}. This would allow for a ML estimation of the power spectrum. It is a promising path that has led to some successes already. However, it is a computationally challenging method that comes with difficulties of its own. We will not follow this route in this work but rather focus on alternative approaches.

To circumvent the numerical difficulties associated with the inversion of the full covariance matrix many approximate methods have been developed. One of the most important approaches is the so-called pseudo-$C_l$ (PCL) approach \cite{WandeltHivonGorski:PCLmethod, Hivon:Master, Efstathiou:MythsandTruths, Hinshaw:FirstYearWMAP} that is discussed in section \ref{subsec:PCL}. While PCL estimation can give rise to good estimators, it is in general not as accurate as the full ML evaluation. The goal of the work discussed in this article is to develop and ultimately implement alternative estimators that improve on PCL estimation and are closer in accuracy to ML estimators. The big advantage of PCL estimation is that the data is compressed into a very limited number of observables which reduces the computational effort tremendously. Unfortunately, the compression is not lossless as soon as statistical anisotropies are introduced. We are going to show how one can improve significantly on PCL estimation by introducing a small number of additional observables that recover the off-diagonal information not picked up in a PCL analysis.

The approach is inspired by the basic idea discussed in \cite{FergussonShellard:OptEst}. Introducing a small statistic quadratic in the multipole coefficients, the authors suggest basing the CMB analysis on this statistic rather than the full set of multipole coefficients. This is of course exactly in the spirit of PCL estimation. However the authors do not restrict themselves to a PCL statistic but rather keep the freedom to choose observables or basis functions in multipole space (giving rise to observables by considering inner products with the data) adapted to the problem at hand. This is motivated by previous work on higher order correlation functions where these modal methods have led to significant numerical simplifications\footnote{In the context of higher order correlation functions the main goal was to arrive at separable approximations to theoretical polyspectra by expanding in a small set of separable modes to reduce the computational complexity of the problem. While separability of the basis functions will speed up calculations in this paper too we will focus on the basic claim that there exist appropriately chosen sets of basis functions that capture nearly all the information but are still small enough to ensure tractability of the problem.} \cite{FergussonReganShellard:trispectrum,FergussonLiguoriShellard:CMBbispectrum,FergussonLiguoriShellard:modeexp}.   

Our purpose here will be to investigate the structure of the masked covariance matrix and hence to identify the additional modes beyond the PCL observables which capture the `off-diagonal' information required for a near optimal analysis. As is discussed in more detail below we will focus on statistical anisotropies introduced by a simple binary mask that vanishes in the masked regions and equals unity elsewhere and we will not be considering instrumental noise throughout most of the discussion. While these assumptions are not a requirement for the suggested methods, they provide a sensible starting point for the discussion. A further publication \cite{Gruetjen:p2} will address more general masks that allow for smooth transitions. These apodised masks are used current in state-of-the-art CMB analyses to reduce coupling of the spherical harmonics. The inclusion of instrumental noise which is necessarily part of any realistic setting will also be studied in the forthcoming publication.

The paper is organised as follows. In section \ref{sec:ParaEst} we start by discussing the ML approach to cosmological parameter estimation. First we review the simplified case of complete sky coverage to develop some basic ideas and then focus on the case of incomplete sky coverage. We briefly study PCL based CMB analysis and then proceed to discuss our approach to the problem. In particular in section \ref{sec:GenApproach} we introduce an approximation to the exact ML evaluation based on an an arbitrary statistic. This approximated estimator is extremely useful as it reduces to PCL estimation if one chooses the simple PCL statistic but for appropriate sets of observables it is exactly equivalent to ML estimation. In this sense it interpolates between PCL and ML estimation and one can systematically improve on PCL estimation by including new observables. We go on to construct a set of observables that in combination with PCL observables leads to significant improvements in the performance of the estimator in the presence of a mask compared to standard PCL estimation. Section \ref{sec:PropCov} contains a systematic analysis that is intended to develop a deeper understanding of the structure of masked covariance matrices and why the new observables improve PCL estimation. Readers that are not interested in these technical details can skip this section and directly move on to section \ref{sec:ChooseBasis} where the new observables and a summary of our method is presented. Results of numerical investigations validating the performance of the estimator can be found in section \ref{sec:Validation}. We conclude by summarising progress with this methodology to date and providing an outlook for what remains to be done in future work.\\[1em]

\begin{center}
\textbf{Conventions}
\end{center}
To ease notation, pairs of indices $(l_i,m_i)$ will usually be denoted as $(l_i,m_i)\equiv\ind_i$.   For example an isotropic covariance matrix of multipole coefficients will be written as
\begin{equation}
\langle a_{l_1m_1}a_{l_2m_2}\rangle=C_{l_1}\delta_{l_1l_2}\delta_{m_1m_2} \equiv C_{l_1} \delta_{\ind_1\ind_2}=\langle a_{\ind_1}a_{\ind_2}\rangle
\end{equation}
Unless stated otherwise, repeated indices are summed over.   
We use a real set of spherical harmonics $Y^R_{\ind} \equiv Y^R_{lm}$ throughout to keep the multipole coefficient of the temperature fluctuations manifestly real.

\section{\label{sec:ParaEst}Cosmological parameter estimation}
As a starting point for our discussion of CMB analysis we assume a data vector $\vec{a}$ consisting of the multipole coefficients $a_{\ind}$ of the measured (pixelised) map of temperature fluctuation $\Delta T(\hat{n}_i)$, that is, with a discrete set of pixel directions $i = 1,\,2,\, ...\, N_\mathrm{pix}$. These are defined by
\begin{equation}\label{eq:maskalm}
a_{\ind}=\sum_{\hat{n}_i}Y^R_{\ind}(\hat{n}_i)\Delta T(\hat{n}_i)U(\hat{n}_i)
\end{equation}
up to a certain $l_{\mathrm{max}}$ depending on the resolution of the experiment. Here, $U(\hat{n})$ is the mask function that vanishes in the masked regions. The mask function accounts for incomplete sky coverage. This is necessary either because the experiment does not observe the entire celestial sphere, which is usually the case for earthbound missions, or because a fraction of the sky is obscured by foreground contamination as discussed in the introduction. In this work we assume a binary mask function that equals unity in observed regions for simplicity. In general, one could also define pseudo multipole coefficients obtained by replacing the mask function $U(\hat{n})$ by a weight function $W(\hat{n})$ that vanishes in contaminated regions but is not restricted to equal 1 elsewhere. In particular one can choose $W(\hat{n})$ that rise smoothly from 0 to 1 so that ringing in multipole space due to discontinuities in pixel space is reduced. The accuracy of PCL estimation can benefit from this (see for example \cite{Hivon:Master, Hansen:Gabor, Smith:PurePCL, Das:EfficientPSEst, Kim:ForegroundMasking, Planck:2013PSandLikelihood}). In this work we specify a binary mask function $U(\hat{n})$ and attempt to construct efficient estimators based on the pseudo-multipole coefficients \eqref{eq:maskalm}. While some of the arguments we will use only strictly apply for such a pure mask function the major conclusion and results also apply to more general weight functions $W(\hat{n})$ (e.g. apodised masks). A future publication will explore the application of the method to apodised masks \cite{Gruetjen:p2}.   

The definition for the $a_{\ind}$ given in \eqref{eq:maskalm} above obviously assumes that the raw data from the satellite has already been processed and compressed into a pixelised map of temperature fluctuations $\Delta T(\hat{n}_i)$. How to compress the time-ordered data into a map has been extensively discussed elsewhere and is not our concern here\footnote{See for example \cite{Tegmark:losslessmaps,Borrill:MADCAP} for early work and \cite{Poutanen:mapmaking} and references within for more recent developments.}.

The CMB is to a very good approximation Gaussian so the probability density function (PDF) of $\vec{a}$ is entirely determined by the covariance matrix of multipole coefficients $\mat{C}$ i.e. given by $\mathcal{P}\left(\vec{a}\vert \mat{C}\right)$. The central problem in CMB analysis is to estimate a set of parameters $\lbrace \epsilon_{\alpha}\rbrace$ that parametrise the covariance matrix $\mat{C}=\mat{C}(\lbrace \epsilon_{\alpha}\rbrace)$.

A standard approach to this problem is using Maximum Likelihood (ML) estimation. In the framework of ML estimation we arrive at estimates $\hat{\epsilon}_{\alpha}$ by maximising the likelihood function given in terms of the PDF via $\mathcal{L}\left(\lbrace \epsilon_{\alpha}\rbrace\vert\vec{a}\right)=(\mathcal{P}\left(\vec{a}\vert \mat{C}(\lbrace \epsilon_{\alpha}\rbrace)\right)$ with respect to the $\epsilon_{\alpha}$. The estimates are then simply given by $\hat{\epsilon}_{\alpha}=\epsilon_{\alpha}^{ML}$ where $\epsilon_{\alpha}^{ML}$ is the set of parameters that maximises the likelihood. Many ML methods have been discussed in the literature and applied to datasets \cite{TegmarkBunn:bruteforce, Gorski:COBEanalysis, Borrill:MADCAP, BondJaffeKnox:estCMB, Tegmark:losslessmeasure}.
To review the basic ideas involved it is sensible to start by studying the case of complete sky coverage, i.e. we set the mask function $U(\hat{n})=1$ everywhere and the multipole coefficients defined above are the true multipoles of the temperature fluctuations.

\subsection{\label{subsec:AnaComp}CMB analysis on the complete sky}
In the case of complete sky coverage the covariance structure of the multipole coefficients $a_{\ind}$ is very simple. Since we expect the fluctuations to be statistically isotropic we need
\begin{equation}\label{eq:isoCl}
(\mat{C})_{\ind_1\ind_2}=\langle a_{\ind_1}a_{\ind_2}\rangle=C_{l_1}\delta_{\ind_1\ind_2}
\end{equation}
where $C_{l}$ is the power spectrum of the CMB. It is straightforward to see that the Gaussian PDF for the temperature fluctuations
\begin{eqnarray}\label{eq:fullskypdf}
\mathcal{P}\left(\vec{a}\vert \mat{C}\right)&=&\frac{1}{N}\exp{\left(-\frac{1}{2}\left(\vec{a}\right)^T\mat{C}^{-1}\vec{a}\right)}\\
N&=&\sqrt{(2\pi)^{\text{rank}(\mat{C})}\mathrm{det}(\mat{C})}
\end{eqnarray}
can be written in terms of a number of $C_l$ estimators $\hat{C}_l$
\begin{eqnarray}
\mathcal{P}\left(\vec{a}\vert \lbrace C_l\rbrace\right)&=&\frac{1}{N}\exp{\left(-\frac{1}{2}\sum_l (2l+1)\frac{\hat{C}_l}{C_l}\right)}\\
N&=&\sqrt{(2\pi)^{(l_{\mathrm{max}}+1)^2}\prod_l C_l^{(2l+1)}}
\end{eqnarray}
where we have substituted \eqref{eq:isoCl} and  we define  
\begin{equation}
\hat{C}_l=\frac{1}{2l+1}\sum_m a_{lm}^2\,.
\end{equation}
It is easy to show that the $\hat{C}_l$ are ML estimators for the $C_l$. Furthermore they are unbiased $\langle \hat{C}_l\rangle=C_l$ and are efficient estimators as they achieve the Cramer-Rao lower bound\footnote{The Cramer-Rao bound in its simplest form is a lower bound on the variance of an unbiased estimator. For an estimator $\hat{\epsilon}$ estimating a single parameter $\epsilon$ without a bias, i.e. one that satisfies $\langle\hat{\epsilon}\rangle=\epsilon$, it states that we need
\begin{equation}
\mathrm{Var}\left(\hat{\epsilon}\right)=\langle\hat{\epsilon}^2\rangle-\langle\hat{\epsilon}\rangle^2\ge \frac{1}{F(\epsilon)}
\end{equation}
where $F(\epsilon)$ is the Fisher information defined by
\begin{equation}
F(\epsilon)=\langle\left(\frac{\partial \log{\mathcal{P}}}{\partial\epsilon}\right)^2\rangle
\end{equation}
The generalisation to an unbiased multi-parameter estimator $\hat{\epsilon}_{\alpha}$ estimating a set of parameters $\epsilon_{\alpha}$ reads
\begin{equation}
\mathrm{Cov}\left(\hat{\epsilon}_{\alpha},\hat{\epsilon}_{\beta}\right)=\langle\hat{\epsilon}_{\alpha}\hat{\epsilon}_{\beta}\rangle-\langle\hat{\epsilon}_{\alpha}\rangle\langle\hat{\epsilon}_{\beta}\rangle\ge F^{-1}_{\alpha\beta}
\end{equation}
where the Fisher information matrix is given by
\begin{equation}
F_{\alpha\beta}=\langle\frac{\partial \log{\mathcal{P}}}{\partial\epsilon_{\alpha}}\frac{\partial \log{\mathcal{P}}}{\partial\epsilon_{\beta}}\rangle
\end{equation}
and the matrix equation $A\ge B$ for two square matrices $A$ and $B$ means that the matrix $A-B$ is positive definite.} on the variance of an unbiased estimator for all values of the parameters $C_l$. In this sense we cannot derive any better estimates for the $C_l$.\\[1em]
Cosmological parameters $\epsilon_{\alpha}$ that determine the power spectrum $C_l(\lbrace\epsilon_{\alpha}\rbrace)$ can be determined by finding their ML values. This can be done using Newton-Raphson iterations giving rise to a Newton-Raphson ML (NRML) estimator. A detailed derivation in the most general case, i.e. not assuming complete sky coverage, can be found in \cite{BondJaffeKnox:estCMB}. It is also reviewed in the appendix \ref{app:ConstrNRML}. It should be pointed out that the procedure is not exactly the Newton-Raphson method but rather an approximation where each step depends on the data in a quadratic fashion. In fact the approximation should be extremely good for the typical sample size of a high resolution CMB experiment. Rather than worrying about this technicality here we will simply ignore it and refer to the quadratic approximation as the NRML estimator (consistent with \cite{Efstathiou:MythsandTruths}) and discuss this issue in the appendix. To find the ML value we start with a fiducial guess $\epsilon_{\alpha}$ giving rise to a fiducial power spectrum $C_l(\epsilon)$ and calculate $\delta \epsilon_{\alpha}$ via
\begin{equation}
\delta \epsilon_{\alpha}=\frac{1}{2}\mathcal{F}^{-1}_{\alpha\beta}\sum_{l}\frac{\partial C_l}{\partial\epsilon_{\beta}}\frac{(2l+1)}{C_l^2}\left(\hat{C}_{l}-C_l\right)
\end{equation}
where 
\begin{equation}
\mathcal{F}_{\alpha\beta}=\frac{1}{2}\sum_{l}\frac{\partial C_l}{\partial\epsilon_{\alpha}}\frac{(2l+1)}{C_l^2}\frac{\partial C_l}{\partial\epsilon_{\beta}}
\end{equation}
This will converge to the ML values of the $\epsilon_{\alpha}$ given that one does not get stuck in a local minimum.\\[1em]
Assessing the accuracy of the parameter estimation is somewhat trickier than in the $\hat{C}_l$ case. It seems very hard to arrive at an analytical expression for the expectation values and variances as we only have an iterative procedure at hand to arrive at the ML values. To obtain an approximate expression for the covariance matrix let us adopt the following point of view. We assume that the sample size is large enough so that the bias of the ML estimates and $\hat{\epsilon}^{ML}_{\alpha}-\epsilon_{\alpha}$ generally are small and the likelihood sufficiently close to Gaussian so that to a good accuracy we obtain the ML estimates from a single iteration
\begin{equation}
\delta \epsilon_{\alpha}=\mathcal{F}^{-1}_{\alpha\beta}\sum_{l}\frac{\partial C_l}{\partial\epsilon_{\beta}}\frac{(2l+1)}{2C_l^2}\left(\hat{C}_{l}-C_l\right)
\end{equation}
away from the true parameter values $\epsilon_{\alpha}$.
In this case we arrive at an estimate for the variance of our estimator given by
\begin{equation}
\langle \Delta \epsilon^{ML}_{\alpha}\Delta \epsilon^{ML}_{\beta}\rangle=\langle \delta \epsilon_{\alpha}\delta \epsilon_{\beta}\rangle=\mathcal{F}^{-1}_{\alpha\beta}
\end{equation}    
Of course in practice we do not know the true values of the parameters. But since we assume the ML estimates to be generally very close to the true values we might as well just evaluate the above expression using the ML values of the parameters for our set of data. In practice this means using the covariance matrix of the last iteration step as an estimate for the covariance matrix of our ML estimates which is also the route taken in \cite{BondJaffeKnox:estCMB}.

Note that this way of assessing errors somewhat assumes that the infinite sample size limit applies. As the ML estimator is expected to be asymptotically efficient, i.e. it achieves the Cramer-Rao bound for all values of the parameters in this limit, this method of assessing errors should give rise to the conclusion that ML estimation is optimal. Indeed it can be shown that we have $\mathcal{F}_{\alpha\beta}=F_{\alpha\beta}$ where $F_{\alpha\beta}$ is the Fisher matrix so that, in this approximation, ML estimation gets assigned the lowest possible error bars as we expected.

Potential issues with this way of assessing errors could arise if the parameters we want to estimate solely influence the power spectrum at low $l$. In this situation it is questionable to what extent the sample size can be considered large and the likelihood might exhibit significant deviations from Gaussianity. This is because there is only a very limited amount of information on the parameters in the small $l$ modes of a particular CMB realisation. Generally, physical parameters will influence the power spectrum at all scales and hence the information content of a particular CMB realisation should become extremely large as one analyses the fluctuations to small scales.

\subsection{\label{subsec:AnaIncomp}CMB analysis on an incomplete sky}
The most severe idealisation in the analysis above was the assumption that we can recover the true multipole coefficients of the CMB from experiments. Due to foreground contamination real CMB experiments can only measure the temperature fluctuations $\Delta T(\hat{n}_i)$ on part of the sky. Hence instead of the true multipole coefficients the data can only be represented by a vector of masked multipole coefficients $\vec{a}$ coefficients given by
\begin{equation}
a_{\ind}=\sum_{\hat{n}_i}Y^R_{\ind}(\hat{n}_i)\Delta T(\hat{n}_i)U(\hat{n}_i)
\end{equation}
up to a certain $l_{\mathrm{max}}$ depending on the resolution of the experiment. Recall that $U(\hat{n})$ is the mask function that vanishes in the masked regions and equals unity elsewhere.

Even though we are now dealing with masked multipole coefficients the statistics clearly remain Gaussian, so it is still possible to express the PDF in the form of the full sky version \eqref{eq:fullskypdf} but with an appropriately modified covariance matrix. However, complications arise because the new covariance matrix $\mat{C}$ becomes singular (observations of different multipole coefficients are no longer statistically independent) and the simple inversion in \eqref{eq:fullskypdf} is not a well-defined operation. Instead, we have to identify what remains of the column space of $\mat{C}$, that is, the image ${\cal{I}}m(\mat{C})$, and then essentially invert $\mat{C}$ restricted to this subspace.  In other words, we need to find the (unique) pseudoinverse $\mat{C}^+$ that satisfies $\mat{C}\,\mat{C}^+\mat{C} = \mat{C}$, $\mat{C}^+\,\mat{C}\mat{C}^+ = \mat{C}^+$, $\left(\mat{C}\,\mat{C}^+\right)^T = \mat{C}\,\mat{C}^+$ and $\left(\mat{C}^+\,\mat{C}\right)^T = \mat{C}^+\,\mat{C}$. Substituting $\mat{C}^+$ for the inverse of the full sky covariance $\mat{C}^{-1}$ in  \eqref{eq:fullskypdf} will then yield the masked PDF (a more detailed discussion can be found in Appendix \ref{app:EffectMask}).
Let us define the following numbers, 
\begin{eqnarray}\label{eq:rank}
{\cal{N}} &:=& \text{rank}\left(\mat{C}\right)\equiv \text{dim}\left({\cal{I}}m(\mat{C})\right)\,,   ~~\hbox{and}\\
{\cal{K}} &:=& \text{nullity}\left(\mat{C}\right )\equiv\, \text{dim}\left({\cal{K}}er(\mat{C})\right)\nonumber
\end{eqnarray}
with  ${\cal{N}}+{\cal{K}}= (l_{\mathrm{max}}+1)^2$ where ${\cal{K}}er(\mat{C})$ is the kernel or null space of $\mat{C}$.    If we introduce ${\cal{N}}$ vectors  $\lbrace \vec{V}_{n}\rbrace$ that form an orthonormal basis of ${\cal{I}}m(\mat{C})$, then any masked realisation of the CMB can be expanded as 
\begin{equation}
\vec{a}=\sum_nA_n \vec{V}_n\quad\hbox{with}\quad A_n=\vec{V}_{n}\cdot \vec{a}\,,
\end{equation}
for an inner product summed over all $l, m$. The PDF for the vector $\vec{A}$ is then
\begin{equation}
\mathcal{P}\left(\vec{A}\vert \mat{C}\right)=\frac{1}{N}\exp{\left(-\frac{1}{2}\vec{a}^T\mat{C}^+\vec{a}\right)}
\end{equation}
with normalisation factor
\begin{equation}
N=\sqrt{(2\pi)^{{\cal N}}\mathrm{det}^+(\mat{C})}\,.
\end{equation}
Here $\mathrm{det}^+(\mat{C})$ is the pseudodeterminant of $\mat{C}$ given by the product of all $\cal N$ non-zero eigenvalues.

We would like to estimate cosmological parameters from this PDF. The construction of the iterative NRML estimator works just as in the full-sky case considered above with the result (compare \cite{BondJaffeKnox:estCMB})
\begin{equation}
\delta \epsilon_{\alpha}=\frac{1}{2}\mathcal{F}^{-1}_{\alpha\beta}\frac{\partial (\mat{C})_{\ind_1\ind_2}}{\partial\epsilon_{\beta}}(\mat{C}^+)_{\ind_1\ind_1^{\prime}}(\mat{C}^+)_{\ind_2\ind_2^{\prime}}\left(a_{\ind_1^{\prime}}a_{\ind_2^{\prime}}-(\mat{C})_{\ind_1^{\prime}\ind_2^{\prime}}\right)
\end{equation}
where
\begin{equation}
\mathcal{F}_{\alpha\beta}=\frac{1}{2}\frac{\partial (\mat{C})_{\ind_1\ind_2}}{\partial\epsilon_{\alpha}}(\mat{C}^+)_{\ind_1\ind_1^{\prime}}(\mat{C}^+)_{\ind_2\ind_2^{\prime}}\frac{\partial (\mat{C})_{\ind_1^{\prime}\ind_2^{\prime}}}{\partial\epsilon_{\beta}}
\end{equation}
Again we refer to the appendix \ref{app:ConstrNRML} for the details of the derivation. Within the approximations discussed in the last section the ML estimator is efficient as it has a variance approximately given by $\mathcal{F}^{-1}_{\alpha\beta}$ and one can show that the Fisher matrix is given by $F_{\alpha\beta}=\mathcal{F}_{\alpha\beta}$.

Unfortunately, direct evaluation of the NRML estimator is not feasible already for moderately large $l_{\mathrm{max}}$ of order $10^2$, as it requires the inversion of the full multipole covariance matrix\footnote{As mentioned in section \ref{sec:Intro} conjugate gradient methods can potentially be employed to evaluate expressions of this type rather than attempting to directly invert the covariance matrix. We will not study this possibility further here}. In the case of the full sky analysis the covariance matrix was simply diagonal but the coupling of the spherical harmonics on an incomplete sky turns the covariance matrix into a complicated dense matrix as is discussed in the next section and inversion is no longer feasible.

\subsection{\label{subsec:MaskCov}Masked covariance matrices}
When we introduce a mask in pixel space we simply set the signal on the contaminated pixels to zero. We can think of this operation as a projection onto the masked sky. In pixel space the operation is extremely simple. However, in multipole space the corresponding projection operator $\mat{P}$ relating configurations on the complete and the masked sky is given by the coupling matrix of spherical harmonics on the masked sky
\begin{equation}\label{eq:projection}
(\mat{P})_{\ind_1\ind_2}=\int\mathrm{d}^2\hat{n}\,U(\hat{n})Y^R_{\ind_1}(\hat{n})Y^R_{\ind_2}(\hat{n})
\end{equation}
Note that, working for now up to arbitrarily high $l$, we can make use of the completeness of the spherical harmonics and we have exactly 
\begin{equation}
\mat{P}^2=\mat{P}\qquad \hbox{and} \qquad \mat{P}^T=\mat{P}\,, 
\end{equation}
so that $\mat{P}$ is indeed an orthogonal projection operator.

Starting with an isotropic theory on the complete sky described by a power spectrum $C_l$ the masked covariance (neglecting instrument noise for now) can be calculated to be
\begin{eqnarray}\nonumber
(\mat{C})_{\ind_1\ind_2}&=&\int\mathrm{d}^2\hat{n}_1\mathrm{d}^2\hat{n}_2\,U(\hat{n}_1)U(\hat{n}_2)Y^R_{\ind_1}(\hat{n}_1)Y^R_{\ind_2}(\hat{n}_2)\\\label{eq:covproj}
&\quad&\qquad\times\underbrace{\langle\Delta T(\hat{n}_1)\Delta T(\hat{n}_2)\rangle}_{C_{l_1^{\prime}}\bar\delta_{\ind_1^{\prime}\ind_2^{\prime}}Y^R_{\ind_1^{\prime}}(\hat{n}_1)Y^R_{\ind_2^{\prime}}(\hat{n}_2)}\\\label{eq:covprojection}
&=&(\mat{P})_{\ind_1\ind_1^{\prime}}C_{l_1^{\prime}}(\mat{P})_{\ind_1^{\prime}\ind_2}
\end{eqnarray}
In principle, the sum over $\ind_1^{\prime}$ in eqn~\eqref{eq:covproj} extends to infinite $l^{\prime}$ unless we are dealing with band limited power spectra. Realistic CMB power spectra decay quickly and so extending the sum to higher $l^{\prime}$ will have little effect on lower $l$ correlations.  If we are only interested in the covariance matrix up to a certain (large) $l_{\mathrm{max}}$ because of beam resolution limits or a poor signal-to-noise ratio, then it is reasonable to cut off the sum without introducing significant errors.

\begin{figure}
\centering
\includegraphics[scale=1]{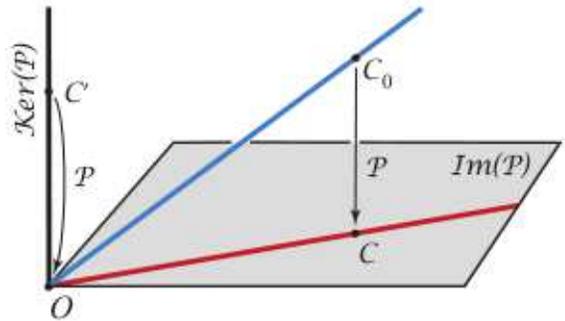}
\caption[Images and kernels of the projection operators]{Schematic of the action of the projection operator $\cal{P}$  induced by masking the sky, see eqns~\eqref{eq:projection} and \eqref{eq:fullcovproj}.  The full sky covariance matrix $\mat{C}_0$ is projected to $\mat{C}$ for the masked sky which lies in the image of $\cal{P}$ (represented by a plane). The subset of isotropic covariance matrices (blue line) is mapped to a subset (red line) lying in ${\cal I}m(\cal{P})$.   The kernel ${\cal{K}}er(\cal{P})$ includes all matrices $\mat{C}'$ which are $\mat{P}$-equivalent to zero.}\label{fig:projection}
\end{figure}

If we include all multipole coefficients in the analysis, notably the monopole and dipole that are not observable in practice\footnote{For consistency we never remove the monopole and dipole in this work from simulated CMB maps but simply assume a value for the power spectrum coefficient $C_0$ and $C_1$. Removing the monopole and dipole, as it has become standard, should not pose significant problems for the methods developed in this work. However, the approach we choose is simpler to understand when the monopole and dipole are included as the interpretation of $\mat{P}$ as a simple orthogonal projection operator is available.} and work up to high $l$, then we can understand the masked covariance matrix as the isotropic diagonal covariance sandwiched between two orthogonal projection operators $\mat{P}$.   This operation 
\begin{equation}\label{eq:fullcovproj}
{\cal P}(\mat{C})=\mat{P}\mat{C}\mat{P}\,,
\end{equation}
inherits its properties from $\mat{P}$ and is itself a projection\footnote{In fact, if we choose the Frobenius inner product as an inner product $\mat{A}:\mat{B}$ on the space of matrices,
\begin{equation}
\mat{A}:\mat{B}=\mathrm{Tr}[\mat{A}\mat{B}^T]\,,
\end{equation}
which is a natural choice in the context of our discussion later on, the projection $\cal{P}$ is an orthogonal projection just like $\mat{P}$. This is easy to see as $\cal{P}$ is clearly self-adjoint with respect to this inner product, $\cal{P}(\mat{A}):\mat{B}=\mat{A}:\cal{P}(\mat{B})$, and a self-adjoint projection is orthogonal. Hence it is sensible to draw image and kernel of $\cal{P}$ at a right angle in figure \ref{fig:projection}.} $\cal{P}$ is  since ${\cal P}\left({\cal P}(\mat{C})\right) =  {\cal P}(\mat{C})$. The action of the projection $\cal P$ is illustrated schematically in fig.~\ref{fig:projection}.  The subset of isotropic covariance matrices $\mat{C}_0$ (represented by a line in fig.~\ref{fig:projection}) is mapped to a set of anisotropic covariance matrices (another line) lying in ${\cal I}m ({\cal P})$.
This picture is useful for the study of estimators on the incomplete sky as in quadratic estimators any matrices weighing the data are generically sandwiched between projected quantities and hence only their projected parts are important.  To formalise this slightly further we introduce the notion of a $\mat{P}$-equivalence relation between matrices induced by the projection $\cal P$. Given a projection operator $\mat{P}$, two matrices $\mat{A}$ and $\mat{B}$ are said to be $\mat{P}$-equivalent if they are the same under the projection $\cal{P}$, that is,
\begin{equation}\label{eq:Pequiv}
\mat{A}~\stackrel{\mat{P}}{\sim}~\mat{B} \qquad \Longleftrightarrow \qquad \mat{P}\mat{A}\mat{P}^T=\mat{P}\mat{B}\mat{P}^T\,
\end{equation}
Clearly, the simplest example is $\mat{ C}_0 = \left(C_{l_1}\delta_{\ind_1\ind_2}\right)\stackrel{\mat{P}}{\sim}\mat{C}$, that is, the isotropic covariance $\mat{ C}_0$  is $\mat{P}$-equivalent to the corresponding masked covariance matrix $\mat{C}$.

Summing up, we see that the masked covariance is significantly different from a simple diagonal matrix due to the non-trivial coupling of spherical harmonics on the masked sky and cannot be inverted analytically. For high resolution experiments the dimension of the covariance matrix is of order $10^7$ and numerical inversion is not  feasible. For this reason, there has been much effort over the recent decades to construct approximate methods that allow a fast estimation of parameters while retaining the accuracy offered by a ML procedure (see for example \cite{WandeltHivonGorski:PCLmethod, Hivon:Master, Efstathiou:MythsandTruths, Hinshaw:FirstYearWMAP}). The most prominent and widely-used method is pseudo-$C_l$ (PCL) estimation. 

\subsection{\label{subsec:PCL}Pseudo-$C_l$ estimation}
Instead of evaluating the full likelihood, PCL estimation relies on the introduction of quantities $\tilde{C}_l$ that are obtained from the $a_{lm}$ just as the $\hat{C}_l$ in the case of complete sky coverage
\begin{equation}
\tilde{C}_l=\frac{1}{2l+1}\sum_m a_{lm}^2
\end{equation}
Using the form of the covariance matrix on a masked sky it is easy to show that they are related to the true power spectrum $C_l$ via
\begin{equation}\label{eq:PCLrelation}
\langle \tilde{C}_{l_1}\rangle=(\mat{M})_{l_1l_2}C_{l_2}
\end{equation}
where\footnote{The sum over $l_3$ in this expression for $\mat{M}$ extends to infinity but in practice one has to cut off the sum at some finite $l_3$. This will introduce errors and lead to slightly biased estimates but usually these effects are small. To avoid any approximations one can calculate $\mat{M}$ directly as is described in appendix \ref{app:Numerics}.}
\begin{equation}
(\mat{M})_{l_1l_2}=(2l_2+1)\sum_{l_3}\frac{2l_3+1}{4\pi}U_{l_3}\left(\begin{array}{ccc} l_1 & l_2 & l_3 \\0&0&0\end{array}\right)^2
\end{equation}
with $U_l$ given in terms of the multipole coeffcients $u_{lm}$ of the mask function $U(\hat{n})$
\begin{equation}
U_l=\frac{1}{2l+1}\sum_m u_{lm}^2
\end{equation}
Given that $\mat{M}$ is invertible, which is usually the case given the fraction $f_{\mathrm{sky}}$ covered by the experiment is not too small (compare \cite{Efstathiou:MythsandTruths} and references within), we can obtain unbiased PCL estimates $\hat{C}^p_l$
\begin{equation}
\hat{C}^p_{l_1}=(\mat{M}^{-1})_{l_1l_2}\tilde{C}_{l_2}
\end{equation}
In contrast to the full-sky case these estimators are not ML estimators for the $C_l$. Furthermore, as is well known, they are in general not optimal. We will discuss this point in more detail below.

A sensible way of estimating cosmological parameters $\epsilon_{\alpha}$  using PCL methods would be, for example, to start with a fiducial guess, calculating the covariance matrix $\langle \Delta \hat{C}^p_{l_1}\Delta \hat{C}^p_{l_2}\rangle$ for the corresponding power spectrum and then minimising the $\chi^2$
\begin{equation}
\chi^2=\left(\hat{C}^p_{l_1}-C_{l_1}\left(\epsilon_{\alpha}\right)\right)\left(\langle \Delta \hat{C}^p\Delta \hat{C}^p\rangle^{-1}\right)_{l_1l_2}\left(\hat{C}^p_{l_2}-C_{l_2}\left(\epsilon_{\alpha}\right)\right)
\end{equation}
We can calculate the next step via a Newton-Raphson iteration for finding the zero of $\partial \chi^2$ given by
\begin{equation}\label{eq:PCLestimator}
\delta \epsilon_{\alpha}=\mathcal{F}^{-1}_{\alpha\beta} \frac{\partial C_{l_1}}{\partial \epsilon_{\beta}}\left(\langle \Delta \hat{C}^p\Delta \hat{C}^p\rangle^{-1}\right)_{l_1l_2}\left(\hat{C}^p_{l_2}-C_{l_2}\right)
\end{equation}
where
\begin{equation}
\mathcal{F}_{\alpha\beta}=\frac{\partial C_{l_1}}{\partial \epsilon_{\alpha}}\left(\langle \Delta \hat{C}^p\Delta \hat{C}^p\rangle^{-1}\right)_{l_1l_2}\frac{\partial C_{l_2}}{\partial \epsilon_{\beta}}
\end{equation}
In the framework of approximations made in the sections above we can use the covariance matrix of the last iteration as an estimate for the covariance matrix, i.e. 
\begin{equation}
\langle \Delta \epsilon_{\alpha}\Delta \epsilon_{\beta}\rangle=\mathcal{F}^{-1}_{\alpha\beta}(\epsilon)
\end{equation}
It has been argued by many authors that PCL estimation is very close to optimal at high $l$ and that there is no real need for a significantly more costly ML estimation. It has been suggested that the known suboptimality of PCL estimation at low $l$ can be circumvented by combining high $l$ PCL estimates with low $l$ ML estimates obtained from smoothed maps \cite{Hivon:Master, Efstathiou:MythsandTruths}. The emerging hybrid estimator is claimed to recover nearly all information on cosmological parameters present in $\vec{a}$, a proposal we will address later on when we study the properties of the masked covariance matrix in a realistic context.   

\section{\label{sec:GenApproach}Generalised approach to CMB parameter estimation}
As discussed above a brute force numerical evaluation of the full NRML estimator
\begin{equation}\label{eq:NRML}
\delta \epsilon_{\alpha}=\frac{1}{2}\mathcal{F}^{-1}_{\alpha\beta}\frac{\partial (\mat{C})_{\ind_1\ind_2}}{\partial\epsilon_{\beta}}(\mat{C}^+)_{\ind_1\ind_1^{\prime}}(\mat{C}^+)_{\ind_2\ind_2^{\prime}}\left(a_{\ind_1^{\prime}}a_{\ind_2^{\prime}}-(\mat{C})_{\ind_1^{\prime}\ind_2^{\prime}}\right)
\end{equation}
is numerically not feasible as it involves the pseudoinverse of the huge covariance matrix. Hence it is desirable to find approximations to the estimator that do not require the inversion of the full covariance matrix but still perform nearly as well. Here we introduce a class of estimators that can be thought of as approximations to the NRML estimator. As special cases this class includes the NRML estimator itself and also the PCL estimator discussed in the section above.

The basic idea is to introduce a set of real basis functions or modes $Q_n(\ind_1,\ind_2)$. The efficacy of the method depends on the overall number $n_\mathrm{max}$ of  modes required  being modest $n_\mathrm{max} \ll (l_{\mathrm{max}}+1)^2$. We can then define a new (iterative) estimator with  (compare \cite{FergussonShellard:OptEst}) 
\begin{eqnarray}\label{eq:ApproxEst}
\delta \epsilon_{\alpha}&=&\frac{1}{2}\Gamma^{-1}_{\alpha\beta}\frac{\partial\vec{\alpha}^T}{\partial\epsilon_{\beta}}\mat{\xi}^{-1}\left(\vec{\beta}-\vec{\alpha}\right)\\\label{eq:GammaApprox}
\Gamma_{\alpha\beta}&=&\frac{1}{2}\frac{\partial\vec{\alpha}^T}{\partial\epsilon_{\alpha}}\mat{\xi}^{-1}\frac{\partial\vec{\alpha}}{\partial\epsilon_{\beta}}
\end{eqnarray}
where the modal coefficients are given by the summations (or inner products)
\begin{eqnarray}\label{eq:modalestimator}
(\vec{\alpha})_n&=&Q_n(\ind_1,\ind_2)\,(\mat{C})_{\ind_1\ind_2}\\\label{eq:modalestimator2}
(\vec{\beta})_{n}&=&Q_{n}(\ind_1,\ind_2\,)a_{\ind_1}a_{\ind_2}\\\label{eq:modalestimator3}
(\mat{\xi})_{n_1n_2}&=&Q_n(\ind_1,\ind_2)\,(\mat{C})_{\ind_1\ind_1^{\prime}}(\mat{C})_{\ind_2\ind_2^{\prime}}\,Q_n(\ind_1^{\prime},\ind_2^{\prime})
\end{eqnarray}
The new modal estimator has a number of desirable properties. Given that the coefficients $\vec{\beta}$ and $\vec{\alpha}$ can be easily calculated this estimator has significant advantages from a computational point of view. The modal covariance matrix $\mat{\xi}$ can then be obtained to arbitrary precision from Monte Carlo sampling via 
\begin{equation}
\mat{\xi}=\frac{1}{2}\langle (\vec{\beta}-\vec{\alpha})(\vec{\beta}-\vec{\alpha})^T\rangle
\end{equation}
and since we are typically dealing with a limited number of basis functions $n_{\mathrm{max}}$ the inversion $\mat{\xi}^{-1}$ becomes a trivial task. The problem reduces to finding a set of basis functions or modes $Q_n(\ind_1,\ind_2)$ with the property that the estimator performs close to optimal. Before proceeding it is useful to discuss a number of special cases. 

\subsection{\label{subsec:OptApprox}NRML estimator and optimal approximations}
Obviously, if we introduce a complete set of basis functions $Q_n$ the estimator is mathematically identical to the full NRML estimator. However, it can also be equivalent to the full estimator even if the modal set is not complete, provided the modal expansion is equivalent to the estimator product \eqref{eq:NRML} up to the projection $\mat{P}$. This can be expressed more formally using the $\mat{P}$-equivalence relation defined earlier in \eqref{eq:Pequiv}, that is, the modal estimator will be optimal if  
\begin{equation}\label{eq:modaloptimality}
\frac{\partial(\mat{C})_{\ind_1\ind_2}}{\partial\epsilon_{\alpha}}(\mat{C}^+)_{\ind_1\ind_1^{\prime}}(\mat{C}^+)_{\ind_2\ind_2^{\prime}}~\stackrel{\mat{P}}\sim~\sum^{n_\mathrm{max}}_n{}\gamma^{\alpha}_n Q_n(\ind_1^{\prime},\ind_2^{\prime})\,.
\end{equation}
The reason for this is because we then have
\begin{eqnarray*}
&&\left(\frac{\partial(\mat{C})_{\ind_1\ind_2}}{\partial\epsilon_{\alpha}}(\mat{C}^+)_{\ind_1\ind_1^{\prime}}(\mat{C}^+)_{\ind_2\ind_2^{\prime}}\right)(\mat{C})_{\ind_1^{\prime}\ind_3}(\mat{C})_{\ind_2^{\prime}\ind_4}Q_m(\ind_3,\ind_4)\\
&&~~=\frac{\partial(\vec{\alpha})_m}{\partial\epsilon_{\alpha}}=\sum \gamma^{\alpha}_n Q_n(\ind_1^{\prime},\ind_2^{\prime})(\mat{C})_{\ind_1^{\prime}\ind_3}(\mat{C})_{\ind_2^{\prime}\ind_4}Q_m(\ind_3,\ind_4)\\
&&~~=\sum \gamma_n^{\alpha} (\mat{\xi})_{nm}\quad\Rightarrow\quad\gamma_n^{\alpha}=\frac{\partial(\vec{\alpha})_m}{\partial\epsilon_{\alpha}}(\mat{\xi}^{-1})_{mn}
\end{eqnarray*}
hence
\begin{eqnarray*}
\mathcal{F}_{\alpha\beta}&=&\frac{1}{2}\frac{\partial (\mat{C})_{\ind_1\ind_2}}{\partial\epsilon_{\alpha}}(\mat{C}^+)_{\ind_1\ind_1^{\prime}}(\mat{C}^+)_{\ind_2\ind_2^{\prime}}\frac{\partial (\mat{C})_{\ind_1^{\prime}\ind_2^{\prime}}}{\partial\epsilon_{\beta}}\\
&=&\frac{1}{2}\gamma_n^{\alpha} Q_n(\ind_1^{\prime}\ind_2^{\prime})\frac{\partial (\mat{C})_{\ind_1^{\prime}\ind_2^{\prime}}}{\partial\epsilon_{\beta}}=\frac{1}{2}\frac{\partial\vec{\alpha}}{\partial\epsilon_{\alpha}}\mat{\xi}^{-1}\frac{\partial\vec{\alpha}}{\partial\epsilon_{\beta}}=\Gamma_{\alpha\beta}
\end{eqnarray*}
and also for the full estimator
\begin{eqnarray*}
\delta \epsilon_{\alpha}&=&\frac{1}{2}\mathcal{F}^{-1}_{\alpha\beta}\frac{\partial (\mat{C})_{\ind_1\ind_2}}{\partial\epsilon_{\beta}}(\mat{C}^+)_{\ind_1\ind_1^{\prime}}(\mat{C}^+)_{\ind_2\ind_2^{\prime}}\left(a_{\ind_1^{\prime}}a_{\ind_2^{\prime}}-(\mat{C})_{\ind_1^{\prime}\ind_2^{\prime}}\right)\\
&=&\frac{1}{2}\Gamma^{-1}_{\alpha\beta}\frac{\partial(\vec{\alpha})_m}{\partial\epsilon_{\alpha}}(\mat{\xi}^{-1})_{mn}Q_n(\ind_1^{\prime}\ind_2^{\prime})\left(a_{\ind_1^{\prime}}a_{\ind_2^{\prime}}-(\mat{C})_{\ind_1^{\prime}\ind_2^{\prime}}\right)\\
&=&\frac{1}{2}\Gamma^{-1}_{\alpha\beta}\frac{\partial\vec{\alpha}}{\partial\epsilon_{\beta}}\mat{\xi}^{-1}\left(\vec{\beta}-\vec{\alpha}\right)
\end{eqnarray*}
We see that it is not at all necessary to introduce a complete set of basis functions. In principle, all that is required is a set of basis functions satisfying \eqref{eq:modaloptimality} and the approximated estimator will be identical to the original NRML estimator.

Note that the condition \eqref{eq:modaloptimality} can also be understood as the requirement that in a vicinity of the true parameter values we can expand the pseudoinverse of the covariance matrix
\begin{equation}
\mat{C}^+_{\ind_1\ind_2}(\epsilon_{\alpha})~\stackrel{\mat{P}}\sim~\sum^{n_\mathrm{max}}_n\kappa^{\alpha}_n(\epsilon_{\alpha}) Q_n(\ind_1,\ind_2)\,.
\end{equation}
Equation \eqref{eq:modaloptimality} then follows with $\gamma^{\alpha}_n=-\partial\kappa^{\alpha}_n/\partial\epsilon_{\alpha}$. The covariance by itself is easier to study so we will usually simply try to approximate $\mat{C}^+$.

Of course we do not know how to calculate $\mat{C}^+$ exactly which is why we are exploring alternative methodologies in the first place.

\subsection{\label{subsec:PCLApprox}Pseudo-$C_l$ as  a modal estimator}
The second case of interest is the choice of PCL basis functions
\begin{equation}
Q_n(\ind_1,\ind_2)=\frac{\delta_{nl_1}\delta_{\ind_1\ind_2}}{2l_1+1}\,,
\end{equation}
for which we then simply have
\begin{eqnarray}
(\vec{\alpha})_{l_1}&=&(\mat{M})_{l_1l_2}C_{l_2}\\
(\vec{\beta})_{l_1}&=&(\mat{M})_{l_1l_2}\hat{C}^p_{l_2}\\
(\mat{\xi})_{l_1l_2}&=&\frac{1}{2}(\mat{M})_{l_1l_1^{\prime}}(\mat{M})_{l_2l_2^{\prime}}\langle\Delta \hat{C}^p_{l_1^{\prime}}\Delta \hat{C}^p_{l_2^{\prime}}\rangle
\end{eqnarray}
Plugging this into the expressions for the approximated NRML modal estimator \eqref{eq:ApproxEst} shows that we arrive exactly at the iterative PCL estimator introduced previously \eqref{eq:PCLestimator}.
 
In the next section we will endeavour to answer the key question about when and why PCL estimation is equivalent, or at least close in accuracy, to the full NRML estimator. The discussion motivates our choice of basis functions in section \ref{sec:ChooseBasis}  and aims at understanding why the extended set of basis functions gives rise to an estimator more accurate than simple PCL estimation.

\section{\label{sec:PropCov}Properties of masked covariance matrices}
\subsection{\label{subsec:ConstPS}Simplest case with constant power spectrum}
Although realistic power spectra are not constant, this case is of some interest because it provides insight into more realistic situations of practical importance.  As we shall see it is the one special case in which the pseudo-$C_l$ estimator proves to be sufficient for achieving optimality.   

For an approximately constant power spectrum $C_l\approx C_0$ the mulitpole covariance matrix is given by $\mat{C}=C_0\mat{P}$ using \eqref{eq:isoCl} and \eqref{eq:covproj}. Since $\mat{P}$ is an orthogonal projection operator it is its own pseudoinverse, i.e. $\mat{P}^+=\mat{P}$ so that $\mat{C}^+=1/C^2_0\mat{P}$. This implies the $\mat{P}$-equivalence,
\begin{equation}
\frac{\partial(\mat{C})_{\ind_1\ind_2}}{\partial\epsilon_{\alpha}}(\mat{C}^+)_{\ind_1\ind_1^{\prime}}(\mat{C}^+)_{\ind_2\ind_2^{\prime}}\propto\frac{\partial(\mat{C})_{\ind_1^{\prime}\ind_2^{\prime}}}{\partial\epsilon_{\alpha}}~\stackrel{\mat{P}}{\sim}~\gamma^{\alpha}_{l_1^{\prime}}\frac{\delta_{\ind_1^{\prime}\ind_2^{\prime}}}{2l_1^{\prime}+1}\,,
\end{equation}
so by the criterion above PCL estimation is manifestly equivalent to the NRML method. This means that if the power spectrum in the data is close to constant and the purpose of the analysis is to estimate slight deviations from the constant shape there is indeed no gain from applying the full NRML procedure. The error bars on any parameters will be just as small using the far less costly PCL analysis.

Now a realistic power spectrum, in particular the concordance $\Lambda$CDM model, is not at all constant. The $l$ dependance at low to moderately high multipoles is a $1/l^2$ drop-off modulated by acoustic peaks. To understand why this complicates the situation we need to study the general structure of covariance matrices in more detail. It will turn out to be useful to study the normalised covariance matrix $\mat{\mathcal{C}}$ defined by 
\begin{equation}\label{eq:NormCov}
(\mat{\mathcal{C}})_{\ind_1\ind_2}=\frac{(\mat{C})_{\ind_1\ind_2}}{C^{\frac{1}{2}}_{l_1}C^{\frac{1}{2}}_{l_2}}\,.
\end{equation}
Understanding this object is just as useful as $\mat{C}$ itself because we have
\begin{equation}
(\mat{C}^+)_{\ind_1\ind_2}\stackrel{\mat{P}}{\sim}\frac{(\mat{\mathcal{C}}^+)_{\ind_1\ind_2}}{C^{\frac{1}{2}}_{l_1}C^{\frac{1}{2}}_{l_2}}
\end{equation}
which is proven in appendix \ref{app:PseudoNorm}. We see that merely pseudoinversion of $\mat{\mathcal{C}}$ is required.

\subsection{\label{subsec:OptProj}General covariance matrices from oblique projections}
In this section we study the structure of masked covariance matrices\footnote{In fact one could also study more general matrices of the type $\mat{A}^{-\frac{1}{2}}\mat{P}\mat{A}\mat{P}\mat{A}^{-\frac{1}{2}}$ for invertible positive definite $\mat{A}$ and any orthogonal projection operator $\mat{P}$ but for isotropic models the more restricted form (with $\mat{A}$ diagonal) provided here is sufficient.} which previously we showed took a simple projected form  \eqref{eq:covproj}. As argued above without loss of generality, we will represent these covariance matrices in a normalised form $\mat{\mathcal{C}}$ defined by:
\begin{eqnarray}
\quad(\mat{\mathcal{C}})_{\ind_1\ind_2}&\equiv&\frac{(\mat{C})_{\ind_1\ind_2}}{C_{l_1}^{\frac{1}{2}}C^{\frac{1}{2}}_{l_2}}
=\frac{1}{C^{\frac{1}{2}}_{l_1}}(\mat{P})_{\ind_1\ind_1^{\prime}}C_{l_1^{\prime}}(\mat{P})_{\ind_1^{\prime}\ind_2}\frac{1}{C^{\frac{1}{2}}_{l_2}}
\end{eqnarray}
The starting point for this general discussion is the insight that we can write
\begin{equation}\label{eq:covdecomp}
\mat{\mathcal{C}}=\mat{P}_C\mat{P}_C^T
\end{equation}
where we normalise the projection operator $\mat{P}$ as 
\begin{equation}\label{eq:projectionC}
(\mat{P}_C)_{\ind_1\ind_2}=\frac{1}{C^{\frac{1}{2}}_{l_1}}(\mat{P})_{\ind_1\ind_2}C^{\frac{1}{2}}_{l_2}
\end{equation}
Like $\mat{P}$ defined in \eqref{eq:projection}, the matrix $\mat{P}_C$ is also a projection operator satisfying $(\mat{P}_C)^2=\mat{P}_C$, $(\mat{P}_C^T)^2=\mat{P}_C^T$.   However, $\mat{P}_C$ is not an orthogonal projection since $\mat{P}_C^T\ne \mat{P}_C$ -- instead it is an oblique projection.   To analyse the structure of $\mat{P}_C$ we consider its four matrix subspaces:  the image (or column space) ${\cal I}m(\mat{P}_C)$, the kernel  (or null space) ${\cal K}er(\mat{P}_C)$, the row space ${\cal I}m(\mat{P}_C^T)$ and the null space  ${\cal K}er(\mat{P}_C^T)$ for the transpose.    Now for any linear mapping, we must have the following orthogonal complements:
\begin{eqnarray}
{\cal{I}}m(\mat{P}_C)&\perp&{\cal{K}}er(\mat{P}_C^T)\,,\\
{\cal{I}}m(\mat{P}_C^T)&\perp&{\cal{K}}er(\mat{P}_C)\,.
\end{eqnarray}
Further, the covariance matrix structure \eqref{eq:covdecomp} implies we also have 
\begin{eqnarray}
{\cal{I}}m(\mat{\mathcal{C}}) = {\cal{I}}m(\mat{P}_C\,\mat{P}_C^T) = {\cal{I}}m(\mat{P}_C)\,,\\
{\cal{K}}er(\mat{\mathcal{C}}) = {\cal{K}}er(\mat{P}_C\,\mat{P}_C^T)= {\cal{K}}er(\mat{P}_C^T)\,.
\end{eqnarray}
Recalling \eqref{eq:rank}, the eigenstructure of $\mat{\mathcal{C}}$ will include  ${\cal N}$ non-zero eigenvalues for the eigenvectors spanning ${\cal{I}}m(\mat{P}_C)$ and ${\cal K}$ zero eigenvalues for those spanning  ${\cal{K}}er(\mat{P}_C^T)$.    Now it can be shown that the non-zero eigenvalues must all have $\lambda_n \ge 1 \;(n= 0,1,2, ..., \,{\cal N})$. This is because an eigenvector $\vec{v}\in{\cal{I}}m(\mat{P}_C)$ with non-zero eigenvalue $\lambda$ must satisfy\footnote{We can define $\mat{P}^T_C\vec{v}-\vec{v}:=\vec{k}$ and we must have $\vec{k}\in{\cal{K}}er(\mat{P}_C^T)$ as $\mat{P}_C^T$ is a projection. This gives $||\mat{P}^T_C\vec{v}||^2=||\vec{k}+\vec{v}||^2=\vec{v}^2+\vec{k}^2+2\vec{v}\cdot\vec{k}$ but since ${\cal{I}}m(\mat{P}_C)\perp{\cal{K}}er(\mat{P}_C^T)$ and $\vec{v}\in{\cal{I}}m(\mat{P}_C)$ we have $\vec{v}\cdot\vec{k}=0$. Thus $||\mat{P}^T_C\vec{v}||^2=\vec{v}^2+\vec{k}^2\ge\vec{v}^2$ or $||\mat{P}^T_C\vec{v}||\ge||\vec{v}||$.} $||\mat{P}^T_C\vec{v}||\ge||\vec{v}||$. Similarly any vector $\vec{u}\in {\cal{I}}m(\mat{P}_C^T)$, in particular $\vec{u}=\mat{P}^T_C\vec{v}$, must satisfy $||\mat{P}_C\vec{u}||\ge||\vec{u}||$. Hence if $\vec{v}$ is a (unit normalised) eigenvector of $\mat{\mathcal{C}}$ with non-zero eigenvalue we need
\begin{equation}
\lambda_v=||\mat{\mathcal{C}}\vec{v}||=||\mat{P}_C\mat{P}^T_C\vec{v}||\ge||\mat{P}^T_C\vec{v}||\ge||\vec{v}||\ge1
\end{equation}
Thus we expect the spectrum of $\mat{\mathcal{C}}$ to consist of ${\cal K}$ zero eigenvalues and ${\cal N}$ eigenvalues that are equal to or larger than unity.

\begin{figure}
\centering
\includegraphics[scale=0.55]{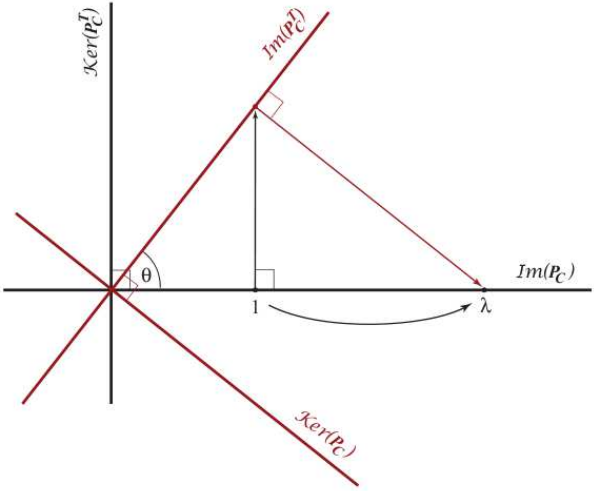}
\caption{Images and kernels of the projection operators in two dimensions for a $\mat{P}$ of rank 1. The successive projections leading to the single nonzero eigenvalue are shown as well.}
\label{fig:2dproj}
\end{figure}

If we restrict attention for a moment to the two-dimensional case with $\mat{P}$ of rank one, we can visualise the situation in a simple two-dimensional context as is shown in  figure \ref{fig:2dproj}.
The kernel of $\mat{\mathcal{C}}$ is just given by ${\cal{K}}er(\mat{P}_C^T)$ and the eigenspace with the single non-zero eigenvalue $\lambda$ is ${\cal{I}}m(\mat{P}_C)$. Note that the two eigenspaces are orthogonal as is required since $\mat{\mathcal{C}}$ is symmetric. The point to emphasise here is that $\lambda$ is exactly unity if ${\cal{I}}m(\mat{P}_C)\perp{\cal{K}}er(\mat{P}_C)$ and can become arbitrarily large when the two are aligned. As discussed above $\lambda$ can never be smaller than one.

Despite the fact that we were discussing the simple two-dimensional case this will still be true in higher dimensions. $\mat{\mathcal{C}}$ will have large eigenvalues when there are directions in ${\cal{I}}m(\mat{P}_C)$ that are almost aligned with directions in ${\cal{K}}er(\mat{P}_C)$. Directions in ${\cal{I}}m(\mat{P}_C)$ that are still close to orthogonal to ${\cal{K}}er(\mat{P}_C)$ will still correspond to eigenvalues close to unity.

This insight is interesting as for masked covariance matrices with power spectra exhibiting strong changes (as it is the case for realistic power spectra) there are generically aligned directions associated with the mask, as we will see below. These directions will come with large eigenvalues that cause the covariance matrix to be different from a simple orthogonal projection operator.

${\cal{I}}m(\mat{P}_C)$ is spanned by the linearly dependent set of unnormalised vectors $\vec{v}_{\ind_0}$ defined by 
\begin{equation}\label{eq:EnhEigen}
(\vec{v}_{\ind_0})_{\ind}=\frac{\mat{P}_{\ind_0\ind}}{C_{l}^{\frac{1}{2}}}=\frac{\left(Y^R_{\ind_0}U\right)_{\ind}}{C_{l}^{\frac{1}{2}}}\,,
\end{equation}
while ${\cal{K}}er(\mat{P}_C)$ is spanned by the linearly dependent set of unnormalised vectors $\vec{k}_{\ind_0}$ given by
\begin{equation}
(\vec{k}_{\ind_0})_{\ind}=\frac{\delta_{\ind_0\ind}-\mat{P}_{\ind_0\ind}}{C_{l}^{\frac{1}{2}}}=\frac{\left(\delta_{\ind_0\ind}-\left(Y^R_{\ind_0}U\right)_{\ind}\right)}{C_{l}^{\frac{1}{2}}}\,.
\end{equation}
Looking at these vectors it is evident that for a decaying power spectrum there must be aligned directions. For a given $\ind$ each pair $\vec{v}_{\ind}$ and $\vec{k}_{\ind}$ are aligned, especially pairs with low $l$. This is due to the decay of the power spectrum that suppresses the difference between the two vectors in the $\ind$ component and upweights their high $l$ components that are identical.

\subsection{\label{subsec:EnhDec}Enhanced eigenvectors from $1/l^2$ power spectra}
As the CMB power spectrum initially decays approximately like $1/l^2$ with some modulations this is an important case to understand. Eventually we will have to take into account that due to beam smearing and Silk damping there will be a much faster exponential decay at high multipoles. This will cause some complications that will be discussed below.\\[1em]
The mask leads to a coupling of different multipoles. To arrive at a simple model for the coupling we can simply assume that we have as a rough order of magnitude estimate $\vert\left(Y^R_{\ind_0}U\right)_{\ind}\vert\propto 1/(\vert l_0-l\vert+1)$. This in turn means that we have for the vectors normalised by $C^{1/2}_l\propto 1/l$
\begin{equation}
\vert(\vec{v}_{\ind_0})_{\ind}\vert,\,\vert(\vec{k}_{\ind_0})_{\ind}\vert\propto l/(\vert l_0-l\vert+1)
\end{equation}
In other words, for moderately large $l_0$ these vectors are strongly peaked at $l_0$ and have most of their support in a vicinity of $l_0$. On the other hand, for low $l_0$ their magnitude is approximately constant and they have significant support for all $l$. To calculate inner products involving  $\vec{v}_{\ind_0}$ or $\vec{k}_{\ind_0}$ with large $l_0$ we can restrict the sum to a vicinity of $l_0$. But for large $l_0$ the power spectrum does not change much for a $1/l^2$ decay and is locally almost constant. This means that the $\vec{v}_{\ind_0}$  with large $l_0$ remain almost orthogonal to the kernel spanned by the $\vec{k}_{\ind_0}$ because any $\left(Y^R_{\ind_0}U\right)_{\ind}$ is exactly orthogonal to all the $\delta_{\ind_0\ind}-\left(Y^R_{\ind_0}U\right)_{\ind}$.

Only the low $l_0$ $\vec{v}_{\ind_0}$ get aligned strongly with the corresponding $\vec{k}_{\ind_0}$ and potentially with each other. Thus any enhanced direction $\vec{v}^C$ must be essentially a linear combination of the $\vec{v}_{\ind}$ with low $l$, i.e. can be written as
\begin{equation}
\left(\vec{v}^C\right)_{\ind}\propto\left(\vec{\Delta}\right)_{\ind_0}\left(\vec{v}_{\ind_0}\right)_{\ind}=\frac{\left(U\left(Y^R_{\ind_0}\left(\vec{\Delta}\right)_{\ind_0}\right)\right)_{\ind}}{C_{l}^{\frac{1}{2}}}
\end{equation}
for some vector $\vec{\Delta}$ with components that are only significantly different from 0 for low $l$.

\begin{figure}[htb]
\raggedright
(a)\\
\centering
\includegraphics[scale=.5]{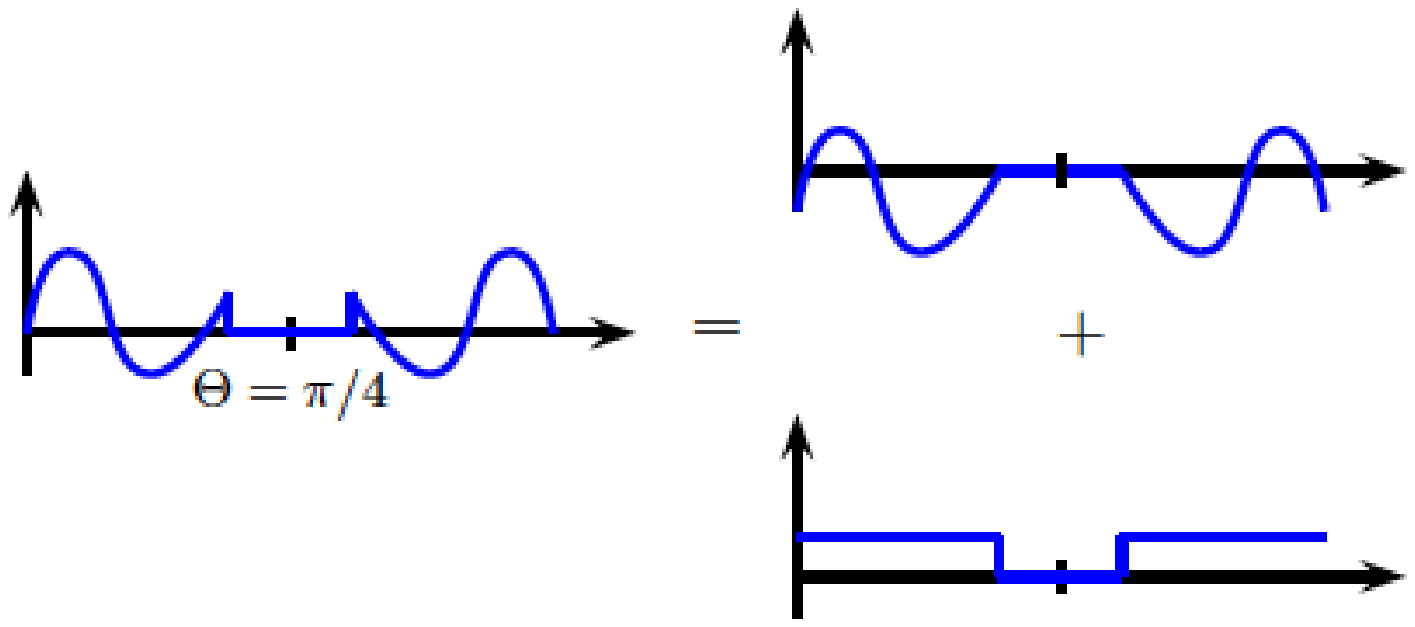}\\
\raggedright
(b)\\
\centering
\includegraphics[scale=0.7]{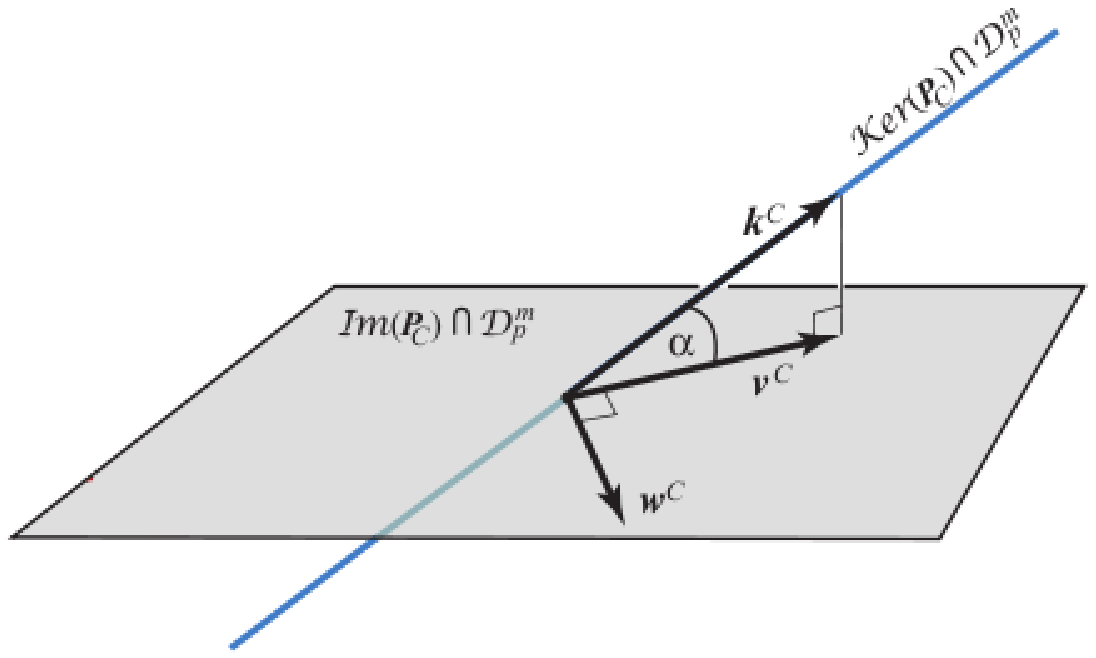}
\caption[(a) Discontinuities in the mask generate high $l$ features in multipole space; generated with PSTricks (b) Schematic picture of image and kernel; generated with PSTricks]{(a) For a pure sky cut symmetric about the equator any of the directions $\vec{v}_{\ind_0}$ for moderately high $l_0$ is essentially the same as $\vec{v}_{\vert m_0\vert m_0}$ or $\vec{v}_{(\vert m_0\vert+1)m_0}$ as the high $l$ behaviour is governed by the discontinuities in $U(\hat{n})Y^R_{\ind_0}(\hat{n})$. The figure shows schematically how these discontinuities (in the $\Theta$ direction) are essentially the same as those of the sky cut. (b) A schematic picture of the image $\mathfrak{I}_C$ and kernel $\mathfrak{K}_C$ of $\mat{P}_C$ restricted to a given $\mathfrak{V}^m_p$, i.e. $\mathfrak{I}_C\cap\mathfrak{V}^m_p$ and $\mathfrak{K}_C\cap\mathfrak{V}^m_p$. All directions in the kernel except for $\vec{k}^C$ are suppressed in this picture to enable us to represent the situation in three dimensions. The directions in the kernel orthogonal to $\vec{k}^C$ should be thought of as being orthogonal to the image as well. The important feature is that any direction in $\mathfrak{I}_C\cap\mathfrak{V}^m_p$ orthogonal to $\vec{v}^C$ is also orthogonal to the kernel and hence cannot correspond to a large eigenvalue.}
\label{fig:eigenvec}
\end{figure}

Further insight can be gained from considering the simplified situation of a pure sky cut that is also symmetric about the equator $\Theta=\pi/4$ instead of the full mask with its asymmetric galactic cut and pointlike features spread over the sky. In this case $\vec{v}_{\ind}$ with different $m$ are mutually orthogonal and the projection operator $\mat{P}_C$ as well as the covariance matrix is block diagonal in $m$. Furthermore, since we assume the sky cut to be symmetric about the equator, we conclude that $\vec{v}_{\ind}$ with even and odd $l$ must be mutually orthogonal due to parity. Similarly $\mat{P}_C$ and $\mat{\mathcal{C}}$ must be block diagonal in parity. This obviously means that we can choose a basis of eigenvectors of $\mat{\mathcal{C}}$ such that every eigenvector is entirely contained in one of the subspaces with a given $m$ and $p$ denoted by $\mathfrak{V}^m_p$ ($p=+$ for even parity and $p=-$ for odd parity). The point is that the set of directions $\vec{v}^C$ parametrised by $\vec{\Delta}$ contains essentially only one vector in each of the $\mathfrak{V}^m_p$. To see this, note that for low $l$ dominated $\vec{\Delta}$ the high $l$ components of $\vec{v}^C$ are governed by the discontinuities in $U(\hat{n})\left(Y^R_{\ind_0}(\hat{n})\left(\vec{\Delta}\right)_{\ind_0}\right)$. As is depicted in figure \ref{fig:eigenvec} (a) we can always split this expression into a sum of a part that looks like the mask itself with its step discontinuities and the associated ringing in multipole space and a part that is continuous and does not contribute significantly to the high $l$ behaviour. But this means that for a given $m$, $p$ all the directions $\vec{v}^C$ parametrised by $\vec{\Delta}$ in the equation above are very aligned. Their low $l$ differences arising from the continuous part in figure  \ref{fig:eigenvec} (a) are suppressed by the scaling of the power spectrum. Hence there can only be one large eigenvector in each $\mathfrak{V}^m_p$. This eigenvector will approximately look like either $\vec{v}_{\vert m\vert m}$ if $m$ even and $p=+$ or $m$ odd and $p=-$ or $\vec{v}_{(\vert m\vert+1)m}$ otherwise.
Of course kernel and image are objects of large dimensions so it is somewhat difficult to get a geometrical intuition. However, we can draw the three dimensional schematic picture in figure \ref{fig:eigenvec} (b) suppressing most directions. It shows geometrically how for given $m$ and $p$ and for a pure sky cut there is a single enhanced direction $\vec{v}^C\in\mathfrak{V}^m_p$ aligned with ${\cal{K}}er(\mat{P}_C)$ and all vectors in ${\cal{I}}m(\mat{P}_C)\cap\mathfrak{V}^m_p$ orthogonal to this vector are also approximately orthogonal to ${\cal{K}}er(\mat{P}_C)$ and hence cannot be eigenvectors with large eigenvalues.

\begin{figure*}[htb]
\centering
\includegraphics[scale=0.35]{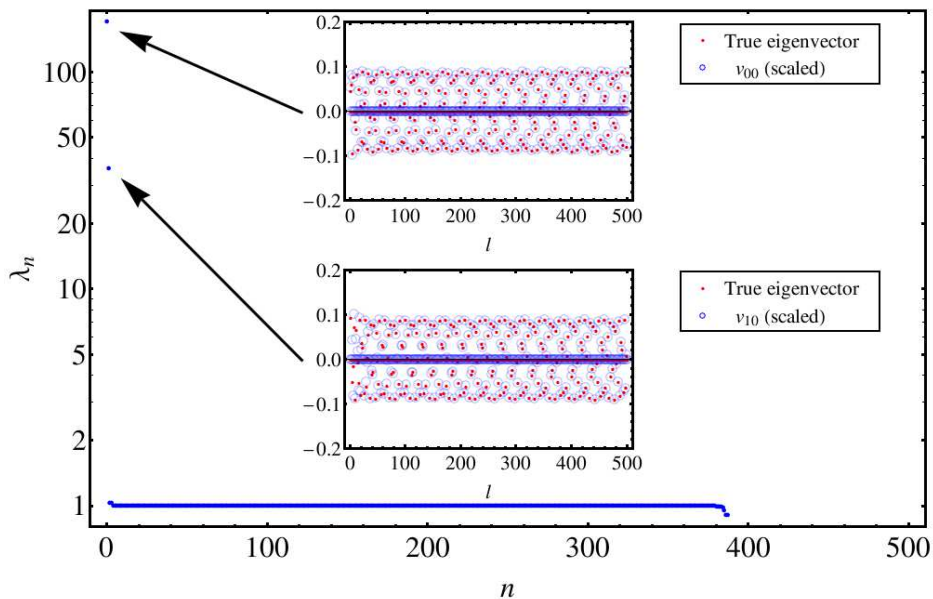}
\caption[The eigenstructure of the $m=0$ block of $\mat{\mathcal{C}}$ for a pure $20^{\circ}$ sky cut and $l_{\mathrm{max}}=500$; plot generated with Mathematica]{The eigenstructure of the $m=0$ block of $\mat{\mathcal{C}}$ for a pure $20^{\circ}$ sky cut and $l_{\mathrm{max}}=500$. The 501 eigenvalues are ordered in decreasing size and labelled by $n$. The insets show the eigenvectors corresponding to the respective two large eigenvalues and compare them to the predictions $\vec{v}_{00}$ and $\vec{v}_{10}$.}
\label{fig:eigenstruc}
\end{figure*}

As the covariance matrix is block diagonal for a pure sky cut it is numerically feasible to calculate its eigensystem. To confirm the ideas above numerically, we picked the $m=0$ block of $\mat{\mathcal{C}}$ and determined its eigenvalues and eigenvectors working up to $l_{\mathrm{max}}=500$ and with a $20^\circ$ cut. Recall that the discussion above suggests that there are merely two enhanced eigenvectors: one corresponding to even parity and the other one to odd parity. The results are shown in figure \ref{fig:eigenstruc} and are clearly consistent with this picture.
The figure shows the eigenstructure of the $m=0$ block of $\mat{\mathcal{C}}$. Among the 501 eigenvalues, ordered in decreasing size and labelled by $n$, there are a number of 0 eigenvalues corresponding to ${\cal{K}}er(\mat{P}_C^T)$ restricted to the $m=0$ subspace\footnote{We made the claim that any non-zero eigenvalue must be $\ge1$. However, we see that the transition from 0 to 1 is not a simple step but there a a few eigenvalues between 0 and 1. This is most likely due to the fact that $\mat{P}$ (and hence $\mat{P}_C$) is not exactly a projection operator unless one works up to infinitely high $l_{\mathrm{max}}$ as we discussed earlier. Thus, it is not inconsistent with the discussion above.}. There are indeed only two large eigenvalues as we would expect: one corresponding to even parity (components with odd $l$ vanish) and the other one to odd parity (components with even $l$ vanish). The insets show the eigenvectors corresponding to the respective two large eigenvalues and compare them to the predictions $\vec{v}_{00}$ and $\vec{v}_{10}$. As expected, they show excellent agreement except for very low $l$. 

If we are not dealing with a pure sky cut and the mask is neither azimuthally symmetric nor parity invariant, we cannot draw the same conclusions on the precise form of the enhanced eigenvectors. The enhanced directions can be any linear combination of the low $l$ $\vec{v}_{\ind}$.
It should be noted that in both cases, given we know the set of enhanced directions and label the corresponding eigenvalues and normalised eigenvectors $\lambda_n$ and $\vec{b}_{n}$, the covariance matrix $\mat{\mathcal{C}}$ can be written as 
\begin{equation}\label{eq:CovApprox}
\mat{\mathcal{C}}\approx\mat{P}^{\perp}_{C}+\sum_n\, \left(\lambda_n-1\right)\vec{b}_{n}\vec{b}_{n}^T
\end{equation}
where $\mat{P}^{\perp}_C$ is the orthogonal projector onto $\mathrm{Im}(\mat{P}_C)$.
We then simply have\footnote{The Moore-Penrose pseudoinverse of $\mat{\mathcal{C}}$ must satisfy $\mat{\mathcal{C}}\mat{\mathcal{C}}^+=\mat{P}^{\perp}_{C}$. It can be seen that this is the case for the expression given above.}
\begin{equation}\label{eq:CovInvApprox}
\mat{\mathcal{C}}^+\approx\mat{P}^{\perp}_{C}-\sum_n\, \left(1-\frac{1}{\lambda_n}\right)\vec{b}_{n}\vec{b}_{n}^T
\end{equation}
Summing up, for the case of a power spectrum that is not constant but approximately decays like $1/l^2$ the masked covariance matrix is different from a simple orthogonal projection operator by a number of enhanced eigenvectors. They are related to masked low $l$ spherical harmonics. In the case of a pure sky cut there are very few of these enhanced eigenvectors due to the high amount of symmetry of the mask.\\[1em]

\subsection{\label{subsec:HighlCorr}The impact of acoustic peaks and beam smearing}
The true situation is more complicated than the scenario discussed above. While the power spectrum decays initially approximately like $1/l^2$, it is modulated by acoustic peaks. At high $l$ power will be exponentially suppressed by Silk damping and on top of that, CMB experiments generically have a finite (albeit small in the case of high resolution experiments) beam width. For a perfectly azimuthally symmetric Gaussian beam, it can be shown that the convolved power spectrum $C_l$ is obtained from the true power spectrum $C^{\mathrm{true}}_l$ via $C_l=b_l^2C^{\mathrm{true}}_l$ where
\begin{equation}
b_l=\exp{\left(-l(l+1)\frac{\sigma^2}{2}\right)}
\end{equation}
and $\sigma=\mathrm{fwhm}/(2\sqrt{2\ln{2}})$ with $\mathrm{fwhm}$ the full width at half maximum of the beam. We see that beam smearing leads to a significant additional exponential suppression of power at high multipoles.

Both of these features lead to changes in the power spectrum and suboptimalities of the PCL estimator. Even though we focused on a simple $1/l^2$ decay above the discussion still essentially applies. The changes in the power spectrum lead to an alignment of image and kernel and result in enhanced eigenvectors of the normalised covariance matrix that need to be accounted for.

Generally these enhanced directions can still be related to masked spherical harmonics with some caveats that will be discussed later on in sections \ref{sec:Validation} and \ref{sec:SummOut}. Of course the changes at high $l$ due to the acoustic peaks and beam suppression will not only align low $l_0$ $\vec{v}_{\ind_0}$ with the kernel. The $\vec{v}_{\ind_0}$ with high $l_0$ can now become aligned with the kernel too.

\section{\label{sec:ChooseBasis}Choosing a near-optimal basis}
Having discussed some features of masked covariance matrices, we will try to exploit the insight we obtained to make a choice of basis functions which gives an approximation to the NRML estimator which improves over PCL estimation. Our strategy will be to start with a set of PCL basis functions\footnote{Note that the normalisation $(2l_1+1)$ in the denominator is not necessary and has no effect on the approximated estimator. However to be more consistent with standard PCL methods we include the normalisation so that the inner products of the basis functions with the data 
\begin{equation}
Q_{(0,n)}(\ind_1,\ind_2)a_{\ind_1}a_{\ind_2}=\sum_m a_{nm}^2/(2n+1)=\tilde{C}_n
\end{equation}
are simply the standard pseudo power spectrum coefficients $\tilde{C}_n$.},
\begin{equation}
Q_{(0,n)}(\ind_1,\ind_2)=\frac{\delta_{nl_1}\delta_{\ind_1\ind_2}}{2l_1+1}\,,
\end{equation}
and complement it with additional basis functions $Q_{(1,n)}$.

\subsection{\label{subsec:AugBasis}Augmented basis beyond PCL}
We saw earlier in section \ref{subsec:OptApprox} that a nearly optimal set of basis function needs to be able to expand $\mat{\mathcal{C}}^+$ in a vicinity of the true values of the parameters. In section \ref{sec:PropCov} we learnt that the normalised covariance $\mat{\mathcal{C}}$ given in \eqref{eq:NormCov} is essentially the orthogonal projection operator $\mat{P}_C^{\perp}$ defined in \ref{subsec:EnhDec} augmented with a number of enhanced eigenvectors, that is, eigenvectors with eigenvalues greater than unity. This yielded a simple expression for $\mat{\mathcal{C}}$ given in \eqref{eq:CovApprox} and a corresponding expression for $\mat{\mathcal{C}}^+$ in \eqref{eq:CovInvApprox}.

PCL basis functions are on their own perfectly able to expand the first part of  $\mat{\mathcal{C}}^+$ arising from $\mat{P}_C^{\perp}$ because of the following equivalence,
\begin{eqnarray}
\frac{1}{C^{\frac{1}{2}}_{l_1}}(\mat{P}_C^{\perp})_{\ind_1\ind_2}\frac{1}{C^{\frac{1}{2}}_{l_2}}&\stackrel{\mat{P}}{\sim}&\frac{1}{C^{\frac{1}{2}}_{l_1}}(\mat{P}_C^{\perp})_{\ind_1\ind_2^{\prime}}\frac{1}{C^{\frac{1}{2}}_{l_2^{\prime}}}(\mat{P})_{\ind_2^{\prime}\ind_2}\\
&=&\frac{1}{C_{l_1}}(\mat{P})_{\ind_1\ind_2}\\
&\stackrel{\mat{P}}{\sim}&\frac{1}{C_{l_1}}\delta_{\ind_1\ind_2}
\end{eqnarray}
In going from the first to the second line we made use of the fact that $(\mat{P})_{\ind_2^{\prime}\ind_2}/C^{{1}/{2}}_{l_2^{\prime}}$ is in the image of $\mat{P}_C$ and thus mapped to itself by the action of $\mat{P}_C^{\perp}$. Hence, we only need to worry about the enhanced directions.

We saw that for a pure sky cut as well as for more general masks\footnote{For a general mask we were still able to conclude that those enhanced eigenvectors associated with the initial $1/l^2$ power law drop-off should be linear combinations of low $l$ $\vec{v}_{\ind}$.} the enhanced eigenvectors of the normalised covariance matrix $\mat{\mathcal{C}}$ are associated with certain 'masked directions' $\vec{v}_{\ind}$ defined in equation \eqref{eq:EnhEigen}.   This motivates the inclusion of  a second set of basis functions
\begin{eqnarray}\label{eq:FullBasis}
Q_{(1,\ind)}(\ind_1,\ind_2)&=&\frac{1}{C^{\frac{1}{2}}_{l_1}}\left(\vec{v}_{\ind}\right)_{\ind_1}\left(\vec{v}_{\ind}\right)_{\ind_2}\frac{1}{C^{\frac{1}{2}}_{l_2}}\\
&=&\frac{1}{C_{l_1}}(\mat{P})_{\ind_1\ind}(\mat{P})_{\ind\ind_2}\frac{1}{C_{l_2}}
\end{eqnarray}
Recall that the index $\ind$ here represents pairs $(l,m)$, so there are a large number of these basis functions. Including all of them up to a certain $l$ means adding another $\left(l+1\right)^2$ basis functions. It is evident that this is only possible up to fairly low $l$ values since otherwise the high number of basis functions renders the problem intractable just as was the case for the original NRML estimator.

A more tractable option, which also gives rise to very accurate estimators, is to choose $m$-summed versions of these basis functions given by
\begin{equation}
Q_{(1,n)}(\ind_1,\ind_2)=\sum_m\frac{1}{C_{l_1}}(\mat{P})_{\ind_1nm}(\mat{P})_{nm\ind_2}\frac{1}{C_{l_2}}
\end{equation}
We find strong evidence that these perform only marginally worse than \eqref{eq:FullBasis} for a $1/l^2$ drop-off of the power spectrum, and lead to significant improvements for the full case when exponential suppression at high $l$ is significant. Their huge advantage is that we can include all of them over the full $l$ range while only doubling the number of basis functions, so that matrix inversion is still tractable.

Furthermore, note that both the $m$-summed and the full set of new basis functions is well suited for numerical evaluation because of their separability and relation to the projection operator. This greatly reduces the numerical effort needed to evaluate the estimator and will be discussed in more detail below.

\subsection{\label{subsec:EffBasis}Demonstrating the efficiency of the augmented estimator}
We proceed by verifying numerically that the new basis functions indeed improve the accuracy of PCL estimators to the extent that they become very nearly optimal. Another key issue is to test whether or not it is sensible to use the $m$-summed set of functions instead of the full set to arrive at approximations to the NRML estimator.

As an example we will consider the case of an amplitude estimator, that is, an estimator that estimates the amplitude of a given fiducial covariance in the data. We choose this estimator because it is particularly simple and it allows for an accurate prediction of the optimal limit. However, it should be noted that this case is also directly relevant for parameter estimation. The overall amplitude of fluctuations is indeed part of the minimal set of parameters that needs to be estimated from the CMB data. We emphasise that including more parameters does not affect the applicability of our method and the same basis will be used later on in the general setting.

An amplitude estimator $\hat{\epsilon}$ estimates the amplitude $\epsilon$ of a covariance $\mat{C}_0$ in the data so that the full covariance is $\mat{C}=\epsilon\,\mat{C}_0$. We then simply have $\mat{C}_{,\epsilon}=\mat{C}_0$. To test our basis functions numerically we used both choices, that is, the full and the $m$-summed set of basis functions combined with the usual PCL basis functions, to arrive at an approximation $\hat{\epsilon}^{\prime}$ to the amplitude estimator $\hat{\epsilon}$ on a sky masked with the WMAP KQ75 mask.

As mentioned above a useful feature of the amplitude estimator is that one can obtain an accurate estimate of what the optimal limit of the variance predicted by the Cramer-Rao bound should be. The ratio of the two is called the efficiency $E$ of the estimator
\begin{equation}
E\left(\hat{\epsilon}^{\prime}\right)=\frac{1/F(\epsilon)}{\mathrm{Var}\left(\hat{\epsilon}^{\prime}\right)}
\end{equation}
Obviously the Cramer-Rao bound requires $E\le 1$ and the larger $E$ the better our approximated estimator uses the information in the data on the parameter of interest. For the true NRML estimator we should have $E=1$, at least in the limit of large sample size. Now for an amplitude estimator,  the Fisher information at $\epsilon=1$ (i.e. for the case that $\mat{\mathcal{C}}_0$ is the true covariance) simplifies to 
\begin{equation}\nonumber
F(\epsilon=1)=\frac{\mathrm{Tr}\left[\left(\mat{C}_0\mat{C}^+_0\right)^2\right]}{2}=\frac{\mathrm{Tr}\left[\mat{P}\right]}{2}=\frac{{\cal{N}}}{2}
\end{equation}
where ${\cal{N}}=\mathrm{rank}(\mat{C}_0)$ as in section \ref{subsec:AnaIncomp}. The discussion of the PDF of multipole coefficients on an incomplete sky in appendix \ref{app:EffectMask} suggests that the rank of a masked covariance matrix should be approximately given by (also compare results in \cite{Challinor:IncSky})
\begin{equation}
{\cal{N}}\approx f_{\mathrm{sky}}(l_{\mathrm{max}}+1)^2
\end{equation}
Hence as a benchmark for our amplitude estimator we estimate the efficiency at $\epsilon=1$ to be
\begin{equation}
E\left(\hat{\epsilon}^{\prime}\right)=\frac{1/F(\epsilon)}{\mathrm{Var}\left(\hat{\epsilon}^{\prime}\right)}=\frac{\frac{\partial\vec{\alpha}^T}{\partial\epsilon}\mat{\xi}^{-1}\frac{\partial\vec{\alpha}}{\partial\epsilon}}{f_{\mathrm{sky}}(l_{\mathrm{max}}+1)^2}
\end{equation}
where $\vec{\alpha}$ and $\mat{\xi}$ are defined in section \ref{sec:GenApproach}. To calculate the variance of the estimator numerically we generate realisation of the CMB and mask them subsequently with the WMAP $KQ75$ mask using HEALPix routines with $N_{\mathrm{side}}=512$. This corresponds to the pixel space resolution of the $KQ75$ mask and should allow for sufficient accuracy working up to $l_{\mathrm{max}}=700$. As the fiducial power spectrum $C_{0;l}$ giving rise to the masked fiducial covariance $\mat{C}_0$ we use the concordance $\Lambda$CDM model. Note that we do not multiply by the beam function initially in order to check the arguments from the previous section for a power spectrum without exponential suppression at high $l$ that causes additional complications. We calculate the vector $\vec{\beta}$ as given in section \ref{sec:GenApproach} for a given number of maps and then obtain $\partial\vec{\alpha}/\partial\epsilon$ and $\mat{\xi}$ via averaging over all maps
\begin{eqnarray}
\frac{\partial\vec{\alpha}}{\partial\epsilon}\Big|_{\epsilon=1}&=&\vec{\alpha}=\langle\vec{\beta}\rangle\\
\mat{\xi}&=&\frac{1}{2}\langle (\vec{\beta}-\vec{\alpha})(\vec{\beta}-\vec{\alpha})^T\rangle
\end{eqnarray}

\begin{figure*}
\centering
\includegraphics[scale=0.35]{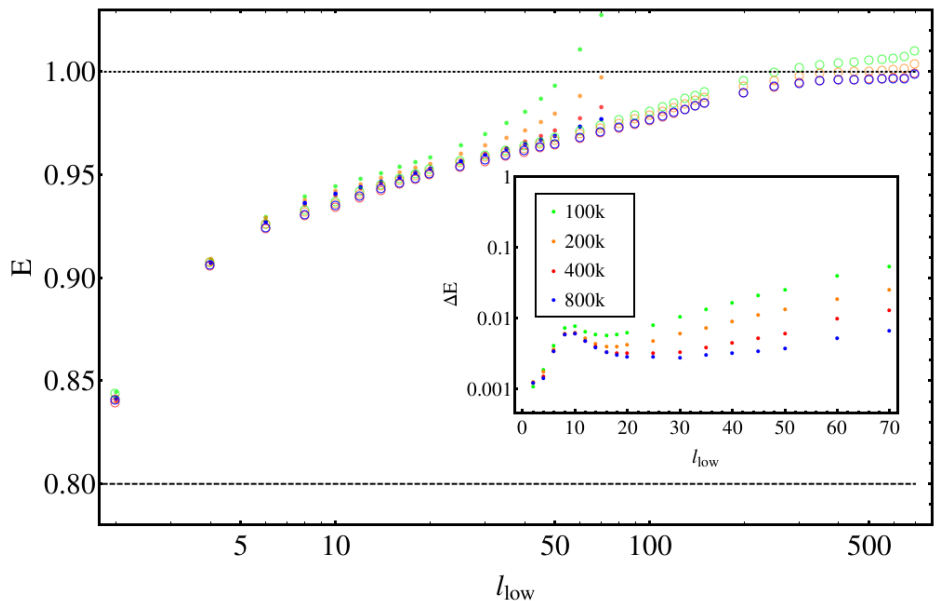}
\caption{The efficiency $E(\epsilon^{\prime})$ of approximations to the NRML amplitude estimator based on PCL basis functions and either the full set of additional functions (dots) or the corresponding $m$-summed versions (circles) up to a given $l_{\mathrm{low}}$. The inset shows the difference in efficiency between the two sets. We obtained both $\vec{\alpha}$ and $\mat{\xi}$ used to calculate $E$ from averaging over MC samples and plotted the curves for averages taken over 100k, 200k, 400k, and 800k MC samples to check convergence. For reference a line with $E=1$ (dotted) corresponding to the optimal limit and a line with $E=0.8$ (dashed) corresponding to the efficiency of PCL estimation without any additional basis functions are added.} 
\label{fig:lowldiff}
\end{figure*}

Figure \ref{fig:lowldiff} is a plot of $E(\epsilon^{\prime})$ for approximations to the NRML estimator using the set of PCL basis functions and either the $m$-summed or the full set of additional basis functions up to a given $l_{\mathrm{low}}$. The inset shows the difference in efficiency between the two sets. For reference we added a line with $E=1$ (dotted) that corresponds to the optimal limit and one with $E=0.8$ (dashed) that corresponds to the efficiency of PCL estimation without any additional basis functions. As we obtained both $\vec{\alpha}$ and $\mat{\xi}$ used to calculate $E$ from averaging over MC samples it is important to check the convergence. We plotted the curves for averages taken over 100k, 200k, 400k, and 800k MC samples. Generally smaller numbers of MC samples lead to an overestimate of the efficiency. For a sufficiently high number of MC samples the curves for the two sets of basis functions are very close to each other and both approach $E\approx 1$ as one increases $l_{\mathrm{low}}$. Hence, using the full set of basis functions only leads to marginal improvements over the $m$-summed set. This is encouraging as it implies that good improvements can in general also be made with the $m$-summed set that remains tractable to arbitrarily high $l_{\mathrm{low}}$. We note that most of the gains in efficiency come from including the lowest $l$ basis functions. This is expected as these should correspond to the large enhanced eigenvectors previously observed in the context of a $1/l^2$ drop-off.

Summing up the analysis of the simple amplitude estimator suggests that strong changes in the power spectrum render PCL analysis suboptimal. Including additional basis functions of the type discussed above helps to recover nearly all of the information lost in a PCL analysis.

As is discussed in \ref{subsec:HighlCorr} exponential suppression at high $l$ due to beam smearing can cause additional severe complications and information loss of PCL estimation. This is effectively a result of the fact that for an unapodised mask there is significant power leakage into high $l$ modes that becomes important relative to the beam suppressed signal. Tests with the amplitude estimator for a power spectrum with beam suppression show that while there are significant improvements in efficiency one does not reach the optimal limit quite as well. We do not include the results here but rather discuss the estimator for a power spectrum with high $l$ suppression from a more general point of view below in section \ref{sec:Validation}.

\subsection{\label{subsec:SummMethod}Summary of the general methodology}
After having introduced the general method of estimation in section \ref{sec:GenApproach} and having chosen a suitable set of basis functions above this is a sensible point to provide a rough summary of our approach for evaluating the iterative estimator. Without any simplifications the steps involved are:

\begin{itemize}
\item[1.] Start with a fiducial guess for the parameters $\epsilon_{\alpha}$. It is sensible to use a fairly advanced guess that is reasonably close to the true parameter values to speed up convergence of the iterative estimator.
\item[2.] Using these parameter values calculate the coefficients $\vec{\alpha}$, $\frac{\partial\vec{\alpha}}{\partial\epsilon_{\alpha}}$ and the covariance matrix of the basis $\mat{\xi}$ defined in \eqref{eq:modalestimator} and \eqref{eq:modalestimator3} respectively assuming the fiducial values of the $\epsilon_{\alpha}$. The basis functions used are the joint set of the standard PCL basis
\begin{equation}\label{eq:PCl}
Q_{(0,n)}(\ind_1,\ind_2)=\frac{\delta_{nl_1}\delta_{\ind_1\ind_2}}{2l_1+1}
\end{equation}
and the new set of basis functions
\begin{equation}\label{eq:augmented}
Q_{(1,n)}(\ind_1,\ind_2)=\sum_m\frac{1}{C_{l_1}}(\mat{P})_{\ind_1nm}(\mat{P})_{nm\ind_2}\frac{1}{C_{l_2}}
\end{equation}
\item[3.] Calculate $\vec{\beta}$ of the actual data. Note that, strictly speaking, this also assumes a choice of fiducial parameters $\epsilon_{\alpha}$ in the definition of the new basis functions.
\item[4.] Calculate $\delta\epsilon_{\alpha}$ via equation \eqref{eq:ApproxEst}.
\item[5.] a) If the calculated adjustment to the fiducial parameters is sufficiently small, i.e. the procedure is converged, then the updated parameter values are the final estimates. An estimate for the errors is given by $\Gamma_{\alpha\beta}^{-1}$ calculated from the quantities obtained in step 2 with the definition of $\Gamma_{\alpha\beta}$ given in equation \eqref{eq:GammaApprox}.\\
b) If the correction to the parameters is significant, the updated parameters are used as the input for step 2 for the next iteration. 
\end{itemize}
The biggest source of computational effort in this procedure is the calculation of $\vec{\alpha}$, $\frac{\partial\vec{\alpha}}{\partial\epsilon_{\alpha}}$ and $\mat{\xi}$ for the chosen set of basis functions in step 2. It is discussed in detail in appendix \ref{app:Numerics}. Here it is enough to note that our new basis functions are chosen such that the computation of the $\beta_n$ is numerically efficient and the calculation of these quantities is feasible. Furthermore one can assume that the initial guess is close enough to the true power spectrum that there is no need to adjust the basis functions or the covariance matrix $\mat{\xi}$ in each step of the iterative procedure. Together with the insight that we have
\begin{equation}
(\vec{\alpha})_{(i,n)}=(\mat{{\cal{M}}})_{(i,n)l}C_l
\end{equation}
this means that each iteration only involves the computation of $C_l(\epsilon_{\alpha})$ and $\frac{\partial C_l}{\partial\epsilon_{\alpha}}$ which is much simpler. $\mat{{\cal{M}}}$ and $\mat{\xi}$ can be obtained as a one-off calculation at the start of the iterative process.

\section{\label{sec:Validation}Validation: covariance matrices of approximated NRML estimators}
So far we have only shown how the new basis functions improve the accuracy of the amplitude estimator as an easy example. As a benchmark for the quality of the approximation we used a single number, namely the efficiency of the estimator. To compare multi-parameter estimators that are used in a realistic analysis we need the full covariance matrices of the parameter estimates.

Suppose we want to estimate a set of parameters $\epsilon_{\alpha}$ with an approximated NRML estimator. The variance of the estimates is then approximately given by $\Gamma^{-1}_{\alpha\beta}$. If we want to compare two approximations $a$, $b$ with covariance matrices $\Gamma^{-1}_{\alpha\beta}(a)$, $\Gamma^{-1}_{\alpha\beta}(b)$ then the natural criterion is that we call the estimator $a$ better than $b$ if as a matrix equation $\Gamma^{-1}_{\alpha\beta}(a)\le\Gamma^{-1}_{\alpha\beta}(b)$ in the sense that $\Gamma^{-1}_{\alpha\beta}(b)-\Gamma^{-1}_{\alpha\beta}(a)$ is positive definite and vice versa. Note that the Cramer-Rao bound requires $\Gamma^{-1}_{\alpha\beta}\ge F_{\alpha\beta}^{-1}$ so there is a limit on how good an estimator can be. $\Gamma^{-1}_{\alpha\beta}(a)\le\Gamma^{-1}_{\alpha\beta}(b)$ is identical to the requirement $\Gamma_{\alpha\beta}(a)\ge\Gamma_{\alpha\beta}(b)$. Now instead of comparing covariance matrices for a large number of conceivable sets of cosmological parameters we can make use of the fact that the covariance depends on the parameters only through the power spectrum coefficients, that is, we can use the chain rule to write \cite{Tegmark:losslessmeasure}
\begin{eqnarray}
\Gamma_{\alpha\beta}&=&\frac{1}{2}\frac{\partial\vec{\alpha}}{\partial\epsilon_{\alpha}}\mat{\xi}^{-1}\frac{\partial\vec{\alpha}}{\partial\epsilon_{\beta}}\\
&=&\frac{1}{2}\frac{\partial C_{l_1}}{\partial \epsilon_{\alpha}}\frac{\partial\vec{\alpha}}{\partial C_{l_1}}\mat{\xi}^{-1}\frac{\partial\vec{\alpha}}{\partial C_{l_2}}\frac{\partial C_{l_2}}{\partial \epsilon_{\beta}}\\
&=&\frac{\partial C_{l_1}}{\partial \epsilon_{\alpha}}\Gamma_{l_1l_2}\frac{\partial C_{l_2}}{\partial \epsilon_{\beta}}\,.
\end{eqnarray}
Hence to make statements about the efficiency of various estimators we can focus on $\Gamma_{l_1l_2}$ as it can be directly related to $\Gamma_{\alpha\beta}$. For example if we have $\Gamma^{-1}_{l_1l_2}(a)\le\Gamma^{-1}_{l_1l_2}(b)$ then we also have $\Gamma^{-1}_{\alpha\beta}(a)\le\Gamma^{-1}_{\alpha\beta}(b)$ for all conceivable parameter estimates so that the approximation a can be considered generally more accurate than b. For this reason we will focus on the $C_l$ covariances $\Gamma^{-1}_{l_1l_2}$ and compare them for different approximations to the NRML estimator.

We will study two cases. First we will work with the WMAP power spectrum without multiplying it by the beam function. This allows us to focus on suboptimalities induced by the initial $1/l^2$ decay and the acoustic peaks. We will then go on to include beam suppression to see how our method performs in this situation.

\subsection{No beam suppression}
\begin{figure*}
\centering
\subfigure[]{\includegraphics[scale=0.25]{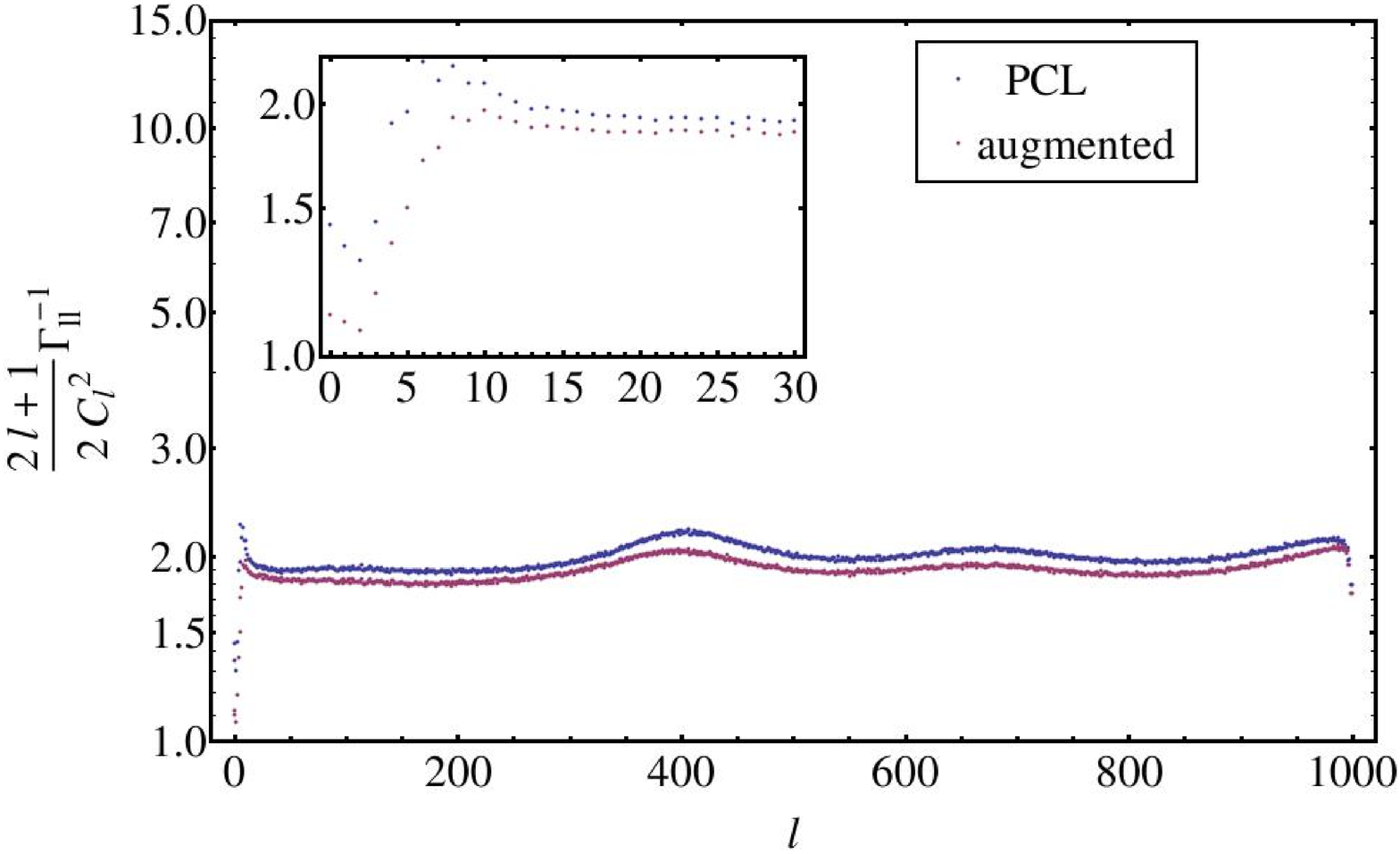}}
\subfigure[]{\raisebox{1.5mm}{\includegraphics[scale=0.245]{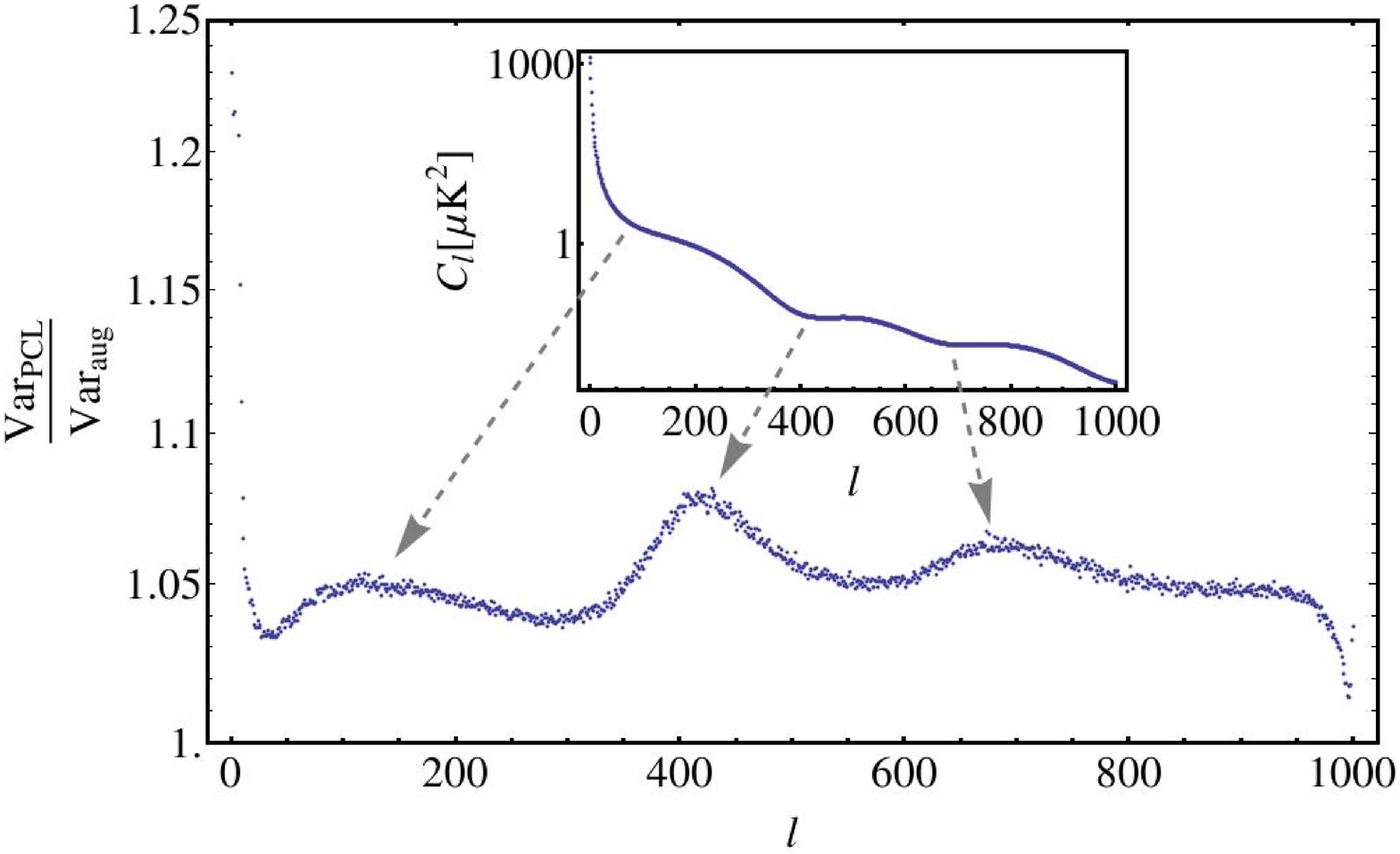}}}
\subfigure[]{\includegraphics[scale=0.4]{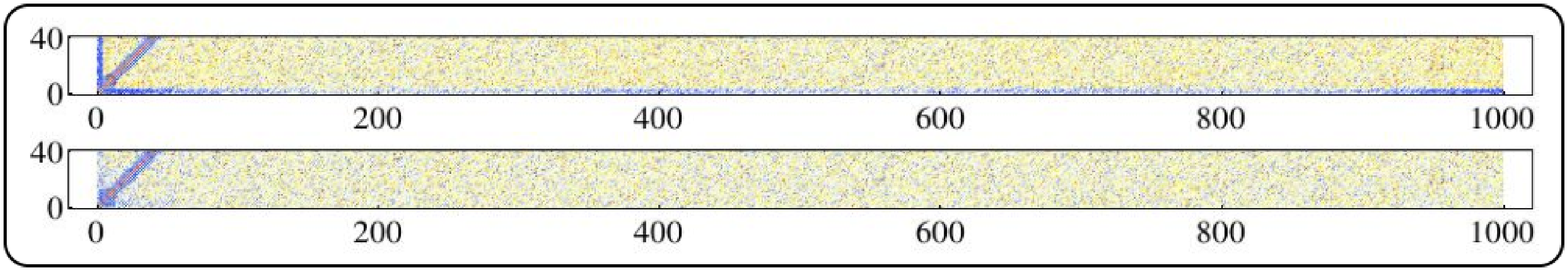}}
\caption{(a) Diagonal entries of the covariances $\Gamma^{-1}_{l_1l_2}$ normalised by full sky variance for PCL basis functions \eqref{eq:PCl} and then augmented with the $m$-summed basis functions \eqref{eq:augmented} for WMAP power spectrum not multiplied by the experimental beam. (b) The ratio of the variances of the PCL estimator and the augmented estimator. The inset shows a log plot of the power spectrum used. ({c}) Density plots showing $40\times1000$ submatrices of the full covariance matrices scaled such that the diagonal entries are exactly unity. The colour scale goes from red (everything above 0.01) to white (0) to blue (everything below -0.01). The upper plot shows the covariance matrix for a PCL basis while the lower plot corresponds to the augmented basis.}
\label{fig:covfullpsnob}
\end{figure*}

Figure \ref{fig:covfullpsnob} depicts the covariance matrices $\Gamma^{-1}_{l_1l_2}$ for different sets of basis functions for the WMAP power spectrum in the absence of beam suppression and without the inclusion of noise. Figure \ref{fig:covfullpsnob} (a) shows the diagonal of $\Gamma^{-1}_{l_1l_2}$ divided by the full sky (cosmic) variance
\begin{equation}
\Delta_{ll}^{\mathrm{full}}=\frac{2\,C_l^2}{2\,l+1}
\end{equation}
for i) a PCL basis and ii) a basis augmented with the new $m$-summed basis functions \eqref{eq:augmented}. The inset highlights the improvements made at low $l$ where the power spectrum changes significantly due to the initial $1/l^2$ decay. Figure \ref{fig:covfullpsnob} (b) provides a plot of the ratio of the variances of the PCL and augmented estimator. Comparing it to the inset showing a log-plot of the power spectrum we observe that the biggest gains are made after drops of the power spectrum either due to the $1/l^2$ scaling at low $l$ or the acoustic peaks at higher $l$. This makes intuitive sense as the contamination due to leakage from lower $l$ modes should be most severe in these spots. Figure \ref{fig:covfullpsnob} ({c}) provides density plots of $40\times1000$ submatrices of the full covariance matrices for the two sets of basis functions. The matrices are scaled such that the diagonal entries are exactly unity, i.e. the plot actually shows $\Gamma^{-1}_{l_1l_2}/(\Gamma^{-1}_{l_1l_1}\Gamma^{-1}_{l_2l_2})^{\frac{1}{2}}$. This makes it easier to plot the off-diagonal elements that are typically much smaller than the diagonal. The top density plot shows the covariance structure for the simple PCL set with off-diagonal low $l$-high $l$ coupling clearly visible. The plot below shows the covariance of an estimator based on PCL functions and $m$-summed basis functions. The plot illustrates how the correlated errors due to the sharp initial $1/l^2$ drop have been largely removed by including the new observables.

The improvements from the augmented basis functions \eqref{eq:augmented} at low $l$ and the mitigation of the low $l$-high $l$ correlations are somewhat similar to what the hybrid estimator mentioned in section \ref{subsec:PCL} is meant to achieve. Instead of combining PCL estimates with further observables constructed from the same map, the hybrid estimator combines them with ML estimates for the low $l$ multipoles obtained from smoothed maps. Whether or not this eliminates the correlated errors in the same way is not entirely clear\footnote{Reference \cite{Efstathiou:MythsandTruths} did not obtain covariance matrices for the hybrid estimator from MC simulations but instead assumed that the ML procedure returns the exact full sky $\hat{C}_l$ estimators and calculated the covariance of the hybrid estimator based on this assumption. This is not necessarily a good assumption because the low $l$ multipoles cannot be exactly recovered for fluctuations that are not band-limited on an incomplete sky.}.

\subsection{With beam suppression}
\begin{figure*}
\subfigure[]{\hspace{-0.05\textwidth}\includegraphics[width=0.6\textwidth]{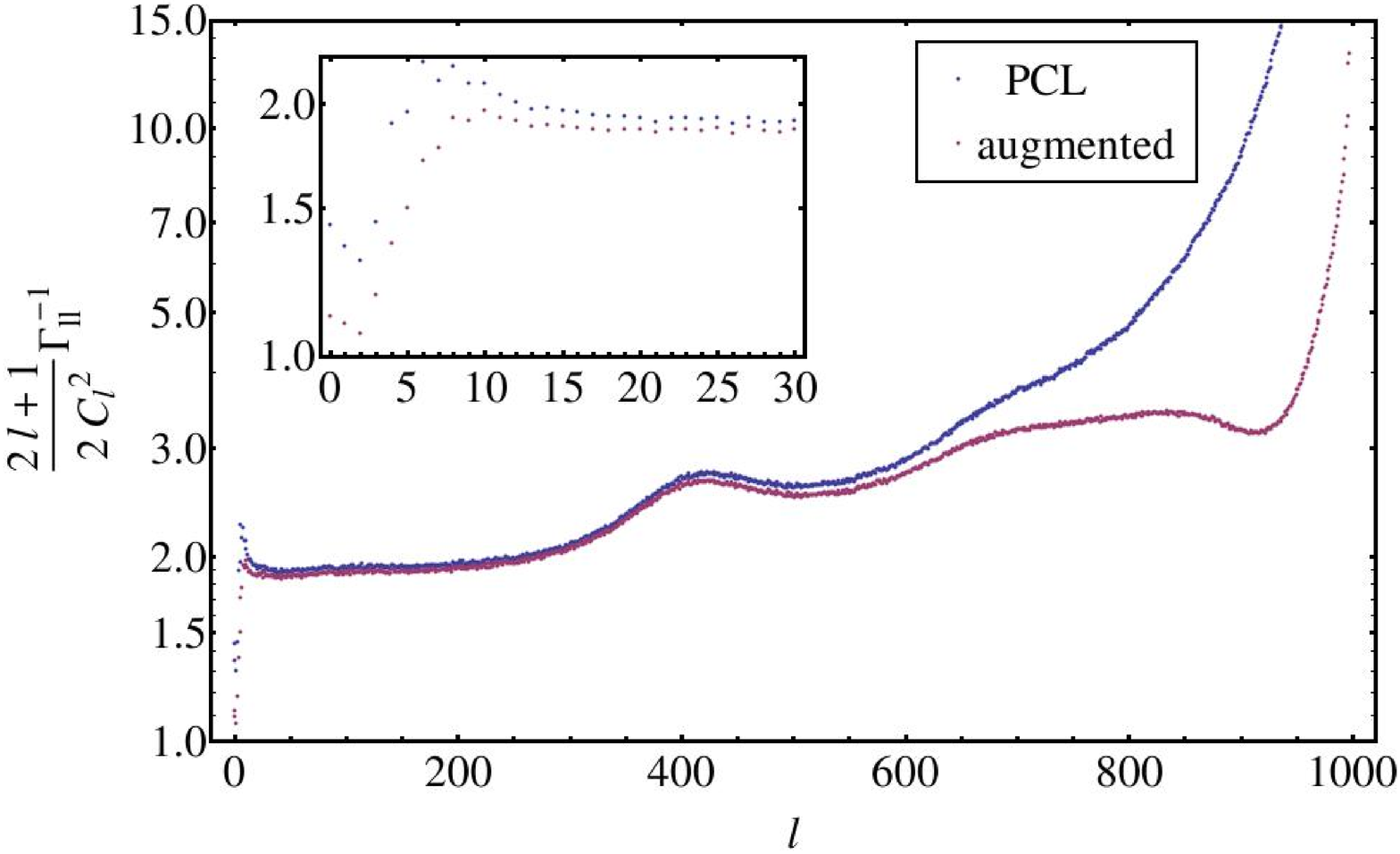}}\\
\subfigure[]{\includegraphics[width=0.7\textwidth]{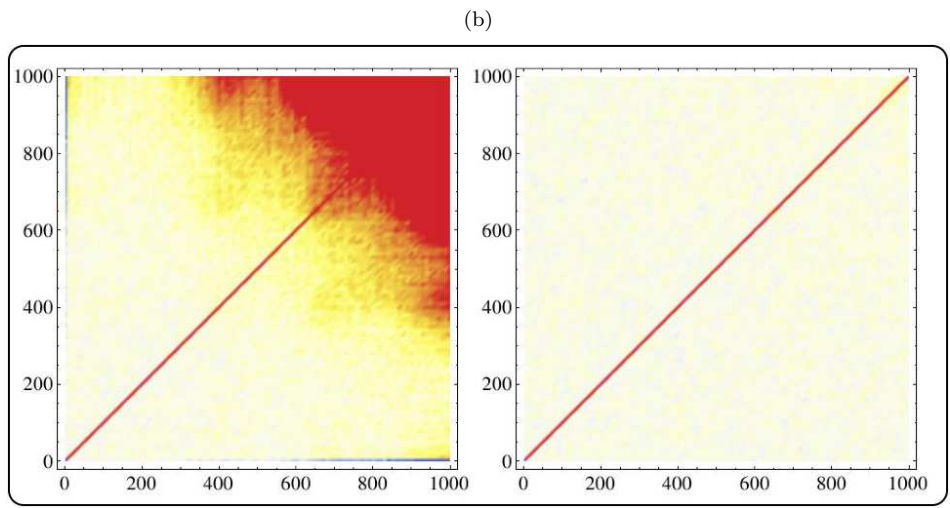}}
\caption{(a) Diagonal entries of the covariances $\Gamma^{-1}_{l_1l_2}$ for PCL basis functions and the augmented basis for WMAP power spectrum multiplied by the experimental beam. (b) Density plots showing the full covariance matrices scaled such that the diagonal entries are exactly unity. The colour scale goes from red (everything above 0.05) to white (0) to blue (everything below -0.05).}
\label{fig:covfullps}
\end{figure*}

Figure \ref{fig:covfullps} is similar to figure \ref{fig:covfullpsnob}, but now the covariance matrices $\Gamma^{-1}_{l_1l_2}$ are based on the WMAP power spectrum multiplied by the WMAP beam. Again figure \ref{fig:covfullps} (a) shows the diagonal of $\Gamma^{-1}_{l_1l_2}$ divided by cosmic variance for the different sets of basis functions. Just as in the previous case without beam suppression the inset emphasises the improvements over PCL that can be made at low $l$. However, we see that the variance of PCL estimates blows up at high $l$ where the power spectrum is exponentially suppressed due to the beam. This is a new feature. It is evident from the plots in figure \ref{fig:covfullps} (a) that significant improvements can be made by including the new basis functions. The augmented basis leads to a drastic reduction in the variance of the estimates at high $l$.

Figure \ref{fig:covfullps} (b) provides density plots showing the off-diagonal structure of the covariance matrices. As in figure \ref{fig:covfullpsnob} the matrices are scaled such that the diagonal elements are unity to enhance the smaller off-diagonal elements. On the left we plotted the covariance resulting from just using PCL basis functions. As in figure \ref{fig:covfullpsnob}, one can see the blue edges corresponding to the low $l$-high $l$ coupling introduced by the strong decay of the power spectrum at low $l$. However, now there is also significant off-diagonal structure at high $l$. To the right is a density plot of the covariance structure for an estimator based on the augmented basis. The inclusion of the new basis functions removes most of the off-diagonal structure both at low and at high $l$, as is shown in the density plot on the right, contrasting starkly with the PCL analysis.

While these results are very encouraging it should be noted that there are several questions that need to be asked at this stage. First of all, while the augmented basis greatly reduces the drastic increase in variance of PCL estimation at high $l$ it does not seem to do a perfect job. The variance is still growing slightly going to higher $l$. The sharp rise near $l_{\mathrm{max}}=1000$ is not unexpected. It is likely due to the fact that the arguments that led us to introduce the augmented basis implicitly assume that there is no cutoff on $l$ in the analysis. Hence, near the cutoff $l_{\mathrm{max}}$ we cannot expect much improvement. However, even before that there is a subtle increase in the variance. This could be interpreted as a hint that the augmented estimator is not recovering all of the information lost in a PCL analysis. Furthermore, comparing the gains in variance to the case without beam suppression it looks like the augmented basis does not lead to a similar improvement after the first acoustic peaks at lower $l$. These observations indicate that the strong leakage due to a binary mask, i.e. the fact that the mask couples multipoles strongly over a long range, not only causes problems for simple PCL estimation in the presence of beam suppression but also diminishes the effectiveness of the augmented basis.

A remedy in this case is apodisation of the mask. By smoothing the edges of a binary mask the long-range coupling of spherical harmonics can be reduced. We will comment on this further in the summary. 

The numerical results in this section where derived using a HEALPix grid with $N_{\mathrm{side}}=512$ which is typically recommended for an analysis up to $l_{\mathrm{max}}\approx 1000$. However, it is clear that when analysing to high multipoles the finite pixelisation starts to play a significant role. In particular one might become concerned that the increase in the variance in the case of beam suppression could be a result of analysis at finite pixel space resolution. This issue is addressed in appendix \ref{app:Pixel} where we note that it is in fact prudent to use pixelisations with $N_{\mathrm{side}}=l_{\mathrm{max}}$ to achieve high accuracy.  However, although the finite resolution has an effect on the variance at high $l$ it is not primarily responsible for the dramatic increase observed in figure \ref{fig:covfullps}.

\section{\label{sec:SummOut}Summary and Outlook}
We have studied the performance of PCL-based CMB analysis on a sky masked with a simple binary mask in the absence of instrumental noise and addressed the question whether PCL estimation can recover information in the data in a nearly optimal fashion, as has been claimed in the literature.
For a power spectrum that exhibits little change over the entire $l$ range, we concluded that PCL estimation does approach optimality. However, this is not the case when large relative changes occur. Realistic power spectra, i.e. power spectra that are measured by high resolution CMB experiments, are far from constant. The concordance $\Lambda$CDM model initially decays like $1/l^2$ and thus exhibits large relative changes at low $l$. Furthermore, relative changes at intermediate $l$ due to acoustic peaks occur and Silk damping and experiment-specific beam smearing eventually lead to an exponential suppression of power which gives rise to a strong decay of the spectrum at high $l$. We have shown that  the decay of the power spectrum leads to the emergence of large eigenvalues in the normalised covariance matrix. Intuitively one can think of these eigenvectors in the covariance matrix as resulting from mode mixing due to the mask. A particular harmonic on the full sky contributes to measurements of other multipole coefficients on the incomplete sky due to the coupling of modes by the mask. If the power spectrum decays quickly, the spurious contribution to the measured signal from modes with smaller $l$ is relatively large compared to the signal at that $l$ value and the pollution is fairly severe. The full NRML estimator discussed above makes use of inverse covariance weighting to simultaneously disentangle the couplings and down-weight the polluted modes in the full dataset and hence gives rise to the lowest possible variance. For a constant power spectrum no down-weighting is necessary and PCL estimation is optimal, but this is not the case in general and PCL basis functions  cannot adequately account for this coupling in the data and do not achieve optimality.

The information loss in the framework of PCL estimation due to a non-constant power spectrum is clearly a phenomenon of practical concern. The important question is whether or not this information loss is small enough to be ignored. If it can be neglected, then PCL analysis is a very attractive method due to its speed and robustness.

The information loss due to the $1/l^2$ drop-off alone (i.e. without further suppression at high $l$ due to a beam function) appears mainly to affect the variance at low $l$ and leads to some low $l$-high $l$ correlation. One could argue that this is not very important because there is not much information in the low $l$ modes. However correlated errors can be important if not taken into account properly. As we have seen for a WMAP mask with an amplitude estimator analysing up to $l_{\mathrm{max}}=700$ without the inclusion of noise, this effect leads to a 20\% increase in the variance due to these correlations of the error bars.

We constructed the augmented basis that is designed to account for the effects of mode coupling on the masked sky. We we were able to recover a very nearly optimal result for the amplitude estimator discussed above.

We proceeded to study the variance of general estimators by focusing on $C_l$ covariance matrices. At higher $l$ acoustic peaks lead to changes of the power spectrum that in turn cause suboptimalities of PCL estimation. We saw how the introduction of the augmented basis improves PCL estimation not only at low $l$, where the $1/l^2$ decay leads to strong changes, but also at higher $l$. Analysing up to $l_{\mathrm{max}}=1000$ for a WMAP power spectrum without including noise and without accounting for beam suppression the augmented basis improves PCL estimation up to 30\% at low $l$ and up to 10\% at higher $l$. The largest gains are made after drops in the power spectrum occur, i.e. after the initial drop and after the locations of the acoustic peaks.

Accounting for power suppression due to beam smearing has a significant impact on the analysis when using a simple binary mask. While the situation at low $l$ is largely unchanged strong decay in the power spectrum at high $l$ introduced by beam smearing leads to very significant effects at high $l$. Again analysing up to $l_{\mathrm{max}}=1000$ without including noise we saw that beam suppression can lead to an increase of the variance of PCL estimation by up to an order of magnitude at high $l$ compared to what one expects from cosmic variance and introduces significant correlations between $C_l$ estimates.

For WMAP these effects set in after the measurements become noise dominated, so it is unlikely that this suboptimality has a large impact on the analysis even when using a binary mask. Here, the noise at high $l$ has already erased most of the information on parameters so there is little to gain. Planck provides an interesting example in which a significantly smaller noise contribution is achieved. If the suppression of the power spectrum due to Silk damping and beam smearing sets in earlier than the point at which the signal becomes noise dominated, then the information loss of PCL estimation becomes more relevant at high $l$.

The augmented basis leads to drastic improvements over PCL estimation, mitigates the increase in the variance and largely removes the strong correlations. However, we saw hints that augmentation alone does not recover all the information. As mentioned above apodisation of the mask which reduces long-range couplings of spherical harmonics is an appropriate measure to tackle the problems due to the suppression of power at high $l$. The Planck analysis made use of this to reduce the information loss of PCL analysis \cite{Planck:2013PSandLikelihood}. While this approach can be successful to avoid an increase of the variance at high $l$ due to long-range power leakage, the coupling of spherical harmonics clearly cannot be avoided completely. Changes of the power spectrum that happen over a smaller $l$ range like the initial $1/l^2$ decay and the acoustic peaks still pose problems. These changes cannot be treated as small over the coupling width of the mask even if we smooth the edges. Furthermore, within the framework of PCL estimation smoothing the edges of the mask is typically accompanied by a loss of effective sky coverage which makes error bars larger at all $l$ so apodising can have negative side effects. We will address these issues in a further publication \cite{Gruetjen:p2}. Our method of augmenting the basis is still available for an apodised mask and is a powerful tool to tackle suboptimalities due to the changes in the power spectrum.

The main development in this work has been to present a general framework in which we can systematically construct approximations to the NRML estimator based on arbitrary sets of basis functions. We have shown how PCL estimation can be incorporated into this framework but, more importantly, how it can be augmented with a small set of projected basis functions \eqref{eq:augmented} chosen to account for the complicated covariance structure of masked multipole coefficients. Including a very moderate number of these new observables can systematically remove suboptimalities of PCL estimation due to a realistic mask. The estimator remains numerically feasible in contrast to naive inverse covariance weighting.

There are a number of open questions and related problems that need to be investigated in future work. In a further publication we will study the effects of the inclusion of noise to determine the significance of the suboptimalities in a more realistic setting. Part of this study is to explore to what extent apodisation of the mask improves PCL estimation at high $l$ and how this smoothing of the mask complements the augmented estimator.

Furthermore a full iterative estimator based on the new $m$-summed modes introduced in this work needs to be implemented to demonstrate feasibility and investigate the potential benefits more thoroughly. In particular, so far we have ignored the effect of the choice of modes on the endpoint of the iterative procedure and simply focused on our estimate for the variance of the estimator as a measure of accuracy.

A natural extension of the method is to construct augmented estimators for the analysis of the CMB polarisation data. A similar approach should lead to significant improvements over the accuracy achieved by simple PCL estimators. This is especially interesting for the estimation of the B-mode which is much smaller in magnitude than the E-mode. Simple PCL estimators mix E and B components and the variance of the B-mode estimates is greatly increased due to leaked E-mode power (see for example \cite{Challinor:ErrorAnaPol}). An augmented estimator could be a promising alternative to pure PCL estimators that have been used to tackle this problem previously \cite{Smith:EBMixing, Smith:GenSolEBMixing}.

The method also has potential applications within the analysis of higher order correlation functions. The search for non-Gaussianity in the CMB fluctuations focuses on the bispectrum, i.e. the 3-point function. Again, the optimal estimator requires inverse covariance weighting and approximate methods related to PCL estimation are usually used to evaluate the estimator. An augmented statistic might help to arrive at nearly optimal estimates.

Of course, the fundamental ideas on how to deal with the effects of masking that we introduced here will also be relevant for observational data in other contexts, such as for the obscuration that occurs in three-dimensional galaxy surveys.

\begin{center}
\textbf{Acknowledgements}
\end{center}
We are very grateful for extensive and valuable discussions with James Fergusson with whom this work is being developed further.   We also thank Marcel Schmittfull for many useful conversations and assistance with computational issues. We are grateful for a critical reading of the present paper by Anthony Challinor and for helpful discussions.  We are also grateful to Andrey Kaliazin for outstanding computational support.  HG gratefully acknowledges the support of the Studienstiftung des deutschen Volkes and an STFC studentship.  This work was supported by STFC grant ST/F002998/1.  Simulations were performed on the COSMOS supercomputer, part of the DiRAC HPC Facility funded by STFC and BIS.   We acknowledge use of the HEALPix package \cite{Gorski:HEALPix}.

\appendix
\section{\label{app:EffectMask}PDF of multipole coefficients on a masked sky}
Masking of certain regions of the sky leads to a singular multipole covariance matrix. As stated above the PDF of multipole coefficients restricted to the subspace ${\cal{I}}m(\mat{C})$ can be written in terms of the unique Moore-Penrose pseudoinverse of the masked covariance matrix $\mat{C}^+$.\\
To understand this in more detail and obtain an intuition let us analyse the situation. Suppose the covariance matrix has a kernel ${\cal{K}}er(\mat{C})$. As the covariance matrix is symmetric ${\cal{K}}er(\mat{C})$ is orthogonal to ${\cal{I}}m(\mat{C})$. Realisations of the CMB on a masked sky $\vec{a}=(a_{\ind})$ must be fully contained in ${\cal{I}}m(\mat{C})$. Hence, if we introduce a set of $\cal{N}=\mathrm{rank}(\mat{C})$ vectors $\lbrace \vec{I}_{n}\rbrace$ that form an orthonormal basis of ${\cal{I}}m(\mat{C})$ any realisation can be expanded as $\vec{a}=A_n \vec{I}_{n}$ with $A_n=\vec{I}_{n}\cdot \vec{a}$. For a more compact notation denote the $(l_{\mathrm{max}}+1)^2\times\cal{N}$ matrix that has the $\vec{I}_n$ as columns by $\mat{I}$. We have 
\begin{equation}
\langle \vec{A} \vec{A}^T\rangle=\mat{I}^T \mat{C}\mat{I}:=\mat{C}_A
\end{equation}
where $\mat{C}_A$ is invertible. We are now in a position to write down the PDF in multipole space for the masked case\footnote{Note that this is now not a density in the full space of harmonic coefficients $\vec{a}$ anymore but rather in the smaller subspace spanned by the $\vec{I}_n$. A density in the full space does not exist for a singular covariance matrix as the Gaussian distribution is only supported on a subset of measure zero.}
\begin{eqnarray}
&\quad&\mathcal{P}\left(\vec{A}\vert \mat{C}\right)\\\nonumber
&=&\frac{1}{N}\exp{\left(-\frac{1}{2}\mat{A}^T\mat{C}^{-1}_A\vec{A}\right)} \\\nonumber
&=&\frac{1}{N}\exp{\left(-\frac{1}{2}\vec{a}^T\left(\mat{I}\mat{C}^{-1}_A\mat{I}^T\right)\vec{a}\right)}
\end{eqnarray}
Now it is true that
\begin{eqnarray}
&\quad&\left(\mat{I}\mat{C}^{-1}_A\mat{I}^T\right)\mat{C}\\
&=&\mat{C}\left(\mat{I}\mat{C}^{-1}_A\mat{I}^T\right)\\
&=&\mat{I}\mat{I}^T:=\mat{P}
\end{eqnarray}
where $\mat{P}$ is the orthogonal projection operator onto ${\cal{I}}m(\mat{C})$. This implies that $\mat{I}\mat{C}^{-1}_A\mat{I}^T$ is the unique Moore-Penrose pseudoinverse\footnote{The Moore-Penrose pseudoinverse of a matrix $\mat{M}$ is defined to be the unique matrix $\mat{M}^+$ satisfying 1) $\mat{M}\mat{M}^+\mat{M}=\mat{M}$, 2) $\mat{M}^+\mat{M}\mat{M}^+=\mat{M}^+$, 3) $\left(\mat{M}\mat{M}^+\right)^{\dagger}=\mat{M}\mat{M}^+$ and 4) $\left(\mat{M}^+\mat{M}\right)^{\dagger}=\mat{M}^+\mat{M}$.} of $\mat{C}$
\begin{equation}
\mat{C}^+=\mat{I}\mat{C}^{-1}_A\mat{I}^T
\end{equation}
We arrive at the final result
\begin{equation}
\mathcal{P}\left(\vec{A}\vert \mat{C}\right)=\frac{1}{N}\exp{\left(-\frac{1}{2}\vec{a}^T\mat{C}^+\vec{a}\right)}
\end{equation}
The normalisation factor is given by
\begin{eqnarray}\nonumber
N&=&\int\mathrm{d}^{\cal{N}}A\exp{\left(-\frac{1}{2}\vec{a}^T\mat{C}^+\vec{a}\right)}\\\nonumber
&=&\int\mathrm{d}^{\cal{N}}A\exp{\left(-\frac{1}{2}\vec{A}^T\mat{C}^{-1}_A\vec{A}\right)}\\\nonumber
&=&\sqrt{(2\pi)^{\cal{N}}\det{(\mat{C}_A)}}\\
&=&\sqrt{(2\pi)^{\cal{N}}\mathrm{det}^*(\mat{C})}
\end{eqnarray}
Where $\mathrm{det}^*$ is the pseudo-determinant (i.e. the product of all non-zero eigenvalues). As required $N$ is like $\mathcal{P}$ as a whole independent of our choice of orthonormal basis in ${\cal{I}}m(\mat{C})$ that was merely an auxiliary construction.\\[1em]
A remaining issue is how to determine the dimension of the image $\cal{N}$. This does not seem to be an easy task in general as the details of the shape of the mask will have to be taken into account. If the $Y^R_{\ind}$ were a complete set of basis functions on the set of $N_{\mathrm{pix}}$ pixels and we leave a set of $N_{\mathrm{pix}}^{\mathrm{unm}}$ pixels unmasked, we would have the exact equality
\begin{equation}
\frac{{\cal{N}}}{(l_{\mathrm{max}}+1)^2}=\frac{N_{\mathrm{pix}}^{\mathrm{unm}}}{N_{\mathrm{pix}}}=f_{\mathrm{sky}}
\end{equation}
where, for sufficiently dense pixelisation, $f_{\mathrm{sky}}$ can be thought of as the fraction of the celestial sphere that is not covered by the mask. This is clear since then we could think of the kernel of $C$ as simply the space of fluctuations that only have support in the masked region. In the realistic case where $(l_{\mathrm{max}}+1)^2<N_{\mathrm{pix}}$ and the $Y^R_{\ind}$ do not form a complete set this is not necessarily true. For reasonably high $N_{\mathrm{pix}}$ and $l_{\mathrm{max}}$ it seems likely that a sensible estimate is given by
\begin{equation}
{\cal{N}}=f_{\mathrm{sky}}(l_{\mathrm{max}}+1)^2
\end{equation}
This is also confirmed by results in \cite{Challinor:IncSky} where amongst other things coupling matrices $(\mat{P})_{\ind_1\ind_2}=\int\mathrm{d}^2\hat{n}Y^R_{\ind_1}Y^R_{\ind_2}U$ and the dimension of their kernels are studied for generic mask functions $U(\hat{n})$.

\section{\label{app:ConstrNRML}Constructing the NRML estimator}
Here we briefly review the construction of the NRML estimator. Most of the details can also be found in \cite{BondJaffeKnox:estCMB}. If we use the Newton-Raphson method to find the zero $\bar{x}$ of a function $f(x)$, i.e. we want to solve $f(x)=0$, we start with a first guess $x_0$ and calculate the next approximation to $\bar{x}$ via
\begin{equation}
x_{n+1}-x_n=\delta_{n+1}=-\frac{f(x_n)}{f^{\prime}(x_n)}
\end{equation}
Given that the procedure will not converge to some other zero this will iterate towards $\bar{x}$. Now if we straightforwardly generalise this procedure to a system of equations and apply it to solve the system
\begin{equation}
\frac{\partial \log{\mathcal{P}}}{\partial \epsilon_{\alpha}}=0
\end{equation}
we arrive at the iterative process
\begin{equation}
\delta\epsilon_{\alpha}=\mathcal{F}^{-1}_{\alpha\beta}\frac{\partial \log{\mathcal{P}}}{\partial\epsilon_{\beta}}
\end{equation}
where
\begin{equation}
\mathcal{F}_{\alpha\beta}=-\frac{\partial^2\log{\mathcal{P}}}{\partial\epsilon_{\alpha}\partial\epsilon_{\beta}}
\end{equation}
Explicitly we obtain
\begin{equation}
\frac{\partial \log{\mathcal{P}}}{\partial\epsilon_{\beta}}=\frac{1}{2}\frac{\partial (\mat{C})_{\ind_1\ind_2}}{\partial\epsilon_{\beta}}(\mat{C}^+)_{\ind_1\ind_1^{\prime}}(\mat{C}^+)_{\ind_2\ind_2^{\prime}}\left(a_{\ind_1^{\prime}}a_{\ind_2^{\prime}}-(\mat{C})_{\ind_1^{\prime}\ind_2^{\prime}}\right)
\end{equation}
and
\begin{equation}
\mathcal{F}_{\alpha\beta}=\mathcal{F}_{\alpha\beta}^0+\delta\mathcal{F}_{\alpha\beta}
\end{equation}
where
\begin{equation}
\mathcal{F}_{\alpha\beta}^0=\frac{1}{2}\frac{\partial (\mat{C})_{\ind_1\ind_2}}{\partial\epsilon_{\alpha}}(\mat{C}^+)_{\ind_1\ind_1^{\prime}}(\mat{C}^+)_{\ind_2\ind_2^{\prime}}\frac{\partial (\mat{C})_{\ind_1^{\prime}\ind_2^{\prime}}}{\partial\epsilon_{\beta}}
\end{equation}
and in more compact notation
\begin{eqnarray*}
\delta\mathcal{F}_{\alpha\beta}&=&\mathrm{Tr}\big[\left(\vec{a}\vec{a}^T-\mat{C}\right)\big(\mat{C}^{-1}\mat{C}_{,\alpha}\mat{C}^{-1}\mat{C}_{,\beta}\mat{C}^{-1}\\
&\quad&\qquad\qquad-\frac{1}{2}\mat{C}^{-1}\mat{C}_{,\alpha\beta}\mat{C}^{-1}\big)\big]
\end{eqnarray*}
Now if we assume that $\mat{C}$ is calculated from parameters that are fairly close to the true parameters then we obviously have $\langle\delta\mathcal{F}_{\alpha\beta}\rangle\approx0$. So $\delta\mathcal{F}_{\alpha\beta}$ is given by a distribution around zero. Fortunately we expect that for a realistic situation this distribution is sharply (compared to the magnitude of $\mathcal{F}_{\alpha\beta}^0$) peaked about its mean zero. For example for the estimation of a single parameter $\epsilon$ that simply changes the amplitude of the covariance, $\mat{C}=\epsilon\mat{C}_0$, we have $\mat{C}_{,\epsilon}=\mat{C}_0$ and $\mat{C}_{,\epsilon\epsilon}=0$ so that we can calculate
\begin{equation}
\langle\delta\mathcal{F}_{\epsilon\epsilon}^2\rangle^{\frac{1}{2}}=2^{\frac{1}{2}}\,\mathrm{rank}\left(\mat{C}_0\right)^{\frac{1}{2}}
\end{equation}
But it is easy to see that $\mathcal{F}_{\epsilon\epsilon}^0=\mathrm{rank}\left(\mat{C}_0\right)/2$ and for a typical high resolution CMB experiment $\mathrm{rank}\left(\mat{C}_0\right)$ is of order $10^6$. This means we expect $\delta\mathcal{F}_{\alpha\beta}$ to be a correction of order $10^{-3}$ to $\mathcal{F}_{\epsilon\epsilon}^0$. Hence in practice we can simply assume
\begin{equation}
\mathcal{F}_{\alpha\beta}\approx\langle\mathcal{F}_{\alpha\beta}\rangle=\mathcal{F}_{\alpha\beta}^0
\end{equation}
The final iterative procedure is thus
\begin{eqnarray*}
\delta\epsilon_{\alpha}&=&\frac{1}{2}\mathcal{F}^{-1}_{\alpha\beta}\frac{\partial (\mat{C})_{\ind_1\ind_2}}{\partial\epsilon_{\beta}}(\mat{C}^+)_{\ind_1\ind_1^{\prime}}(\mat{C}^+)_{\ind_2\ind_2^{\prime}}\left(a_{\ind_1^{\prime}}a_{\ind_2^{\prime}}-(\mat{C})_{\ind_1^{\prime}\ind_2^{\prime}}\right)\\
\mathcal{F}_{\alpha\beta}&=&\frac{1}{2}\frac{\partial (\mat{C})_{\ind_1\ind_2}}{\partial\epsilon_{\alpha}}(\mat{C}^+)_{\ind_1\ind_1^{\prime}}(\mat{C}^+)_{\ind_2\ind_2^{\prime}}\frac{\partial (\mat{C})_{\ind_1^{\prime}\ind_2^{\prime}}}{\partial\epsilon_{\beta}}
\end{eqnarray*}
Although this is, precisely speaking, only an approximation to the Newton-Raphson method for finding the maximum of the likelihood, we refer to this as the Newton-Raphson ML estimator (NRML) estimator. This is consistent with the nomenclature in reference \cite{Efstathiou:MythsandTruths}. We have argued that this is in fact a very good approximation to the actual NRML estimator. However we have seen that the full NRML estimator is not simply quadratic in the data like the approximated version. Reference \cite{BondJaffeKnox:estCMB} therefore calls it the quadratic estimator to emphasise this aspect. We chose not to do this to avoid confusion as, for example, the PCL estimator or any other approximations to the NRML estimator discussed in this paper are quadratic in the data.\\
As is pointed out in \cite{BondJaffeKnox:estCMB} this approximation should have no effect on the endpoint. Both methods should converge to the ML value of the parameters. So at most this will affect the speed of convergence.

\section{\label{app:PseudoNorm}Pseudoinversion of normalised covariance}
We want to show that
\begin{equation}
(\mat{C}^+)_{\ind_1\ind_2}\stackrel{\mat{P}}{\sim}\frac{(\mat{\mathcal{C}}^+)_{\ind_1\ind_2}}{C^{\frac{1}{2}}_{l_1}C^{\frac{1}{2}}_{l_2}}
\end{equation}
Note that this is not a simple matrix identity. An easy way to see this is noting that the two sides do not have the same kernel.

Let us study a slightly more general setting and introduce a normalised covariance given by $\mat{A}^{-1}\mat{C}(\mat{A}^{-1})^T$ for some invertible matrix $\mat{A}$ (in the case of interest just pick $(\mat{A})_{\ind_1\ind_2}=C^{\frac{1}{2}}_{l_1}\delta_{\ind_1\ind_2}$). We are going to show that as a matrix identity
\begin{equation}
\mat{C}^+=\mat{P}(\mat{A}^{-1})^T\left(\mat{A}^{-1}\mat{C}(\mat{A}^{-1})^T\right)^+\mat{A}^{-1}\mat{P}
\end{equation}
where $\mat{P}$ is the orthogonal projection operator onto ${\cal{I}}m(\mat{C})$. Note that for one thing both sides have the same kernel given by the vector space $\lbrace(\mathbb{I}-\mat{P})\vec{v}, \vec{v}\in\mathbb{R}^{(l_{\mathrm{max}}+1)^2}\rbrace$ and the same image given by $\lbrace \mat{P}\vec{v}, \vec{v}\in\mathbb{R}^{(l_{\mathrm{max}}+1)^2}\rbrace$.\\[1em]
Proof:\\
Let $\mat{P}_{A}$ be the orthogonal projection operator onto the subspace $\lbrace \mat{A}^{-1}\mat{P}\vec{v}, \vec{v}\in\mathbb{R}^{(l_{\mathrm{max}}+1)^2}\rbrace$. We must have $\mat{A}^{-1}\mat{P}=\mat{P}_{A}\mat{A}^{-1}\mat{P}$. Using 
\begin{eqnarray*}
\mat{P}&=&\mat{C}\mat{C}^+\\
\mat{P}_{A}&=&\left(\mat{A}^{-1}\mat{C}(\mat{A}^{-1})^T\right)\left(\mat{A}^{-1}\mat{C}(\mat{A}^{-1})^T\right)^+
\end{eqnarray*}
we arrive at
\begin{eqnarray*}
&\quad&\mat{A}^{-1}\mat{C}\mat{C}^+=\mat{A}^{-1}\mat{P}=\mat{P}_{A}\mat{A}^{-1}\mat{P}\\
&=&\left(\mat{A}^{-1}\mat{C}(\mat{A}^{-1})^T\right)\left(\mat{A}^{-1}\mat{C}(\mat{A}^{-1})^T\right)^+\mat{A}^{-1}\mat{P}
\end{eqnarray*}
acting on both sides with $\mat{C}^+\mat{A}$ gives
\begin{eqnarray*}
&\quad&(\mat{C}^+\mat{A})\mat{A}^{-1}\mat{C}\mat{C}^+=\mat{C}^+\\
&=&(\mat{C}^+\mat{A})\left(\mat{A}^{-1}\mat{C}(\mat{A}^{-1})^T\right)\left(\mat{A}^{-1}\mat{C}(\mat{A}^{-1})^T\right)^+\mat{A}^{-1}\mat{P}\\
&=&\mat{P}(\mat{A}^{-1})^T\left(\mat{A}^{-1}\mat{C}(\mat{A}^{-1})^T\right)^+\mat{A}^{-1}\mat{P}
\end{eqnarray*}
Hence we have shown 
\begin{equation}
\mat{C}^+=\mat{P}(\mat{A}^{-1})^T\left(\mat{A}^{-1}\mat{C}(\mat{A}^{-1})^T\right)^+\mat{A}^{-1}\mat{P}
\end{equation}
as intended.\\[1em]
Using $\mat{P}$-equivalence this can be stated as
\begin{equation}
\mat{C}^+\stackrel{\mat{P}}{\sim}(\mat{A}^{-1})^T\left(\mat{A}^{-1}\mat{C}(\mat{A}^{-1})^T\right)^+\mat{A}^{-1}
\end{equation}
which is the desired result for our special case $(\mat{A})_{\ind_1\ind_2}=C^{\frac{1}{2}}_{l_1}\delta_{\ind_1\ind_2}$.

\section{\label{app:Numerics}Calculation of $\beta_n$, $\mat{\xi}$, $\vec{\alpha}$ and $\frac{\partial\vec{\alpha}}{\partial\epsilon_{\alpha}}$}
The calculation of the $\beta_n$ for a given data vector $a_{lm}$ is very fast. Calculating the PCL coefficients $\beta_{(0,n)}=1/(2n+1)\sum_m a_{nm}^2$ is easy as it only involves a summation over $m$. Our choice of new basis functions also allows for efficient evaluation of the inner products with the data. The inner product does not require $\mathcal{O}(l_{\mathrm{max}}^4)$ operations as one would naively expect. One can see that the remaining coefficients are given by
\begin{eqnarray}\nonumber
\beta_{(1,n)}&=&Q_{(1,n)}(\ind_1,\ind_2)a_{\ind_1}a_{\ind_2}\\
&=&\sum_m\left(P_{nm,l_1m_1}\frac{a_{l_1m_1}}{C_{l_1}}\right)^2
\end{eqnarray}
They can be computed from the data by generating a map generated from the $C_l^{-1}$ weighted multipoles $a_{lm}/C_l$ that is subsequently masked and reanalysed (this is the action of $\mat{P}$). The $\beta_{(1,n)}$ are then obtained as PCL-like sums of the new multipoles. Due to the fast HEALPix routines for generating and analysing maps the calculations can be performed rapidly.

Since we have a fast way of obtaining $\vec{\beta}$ for maps we can use MC sampling to calculate $\mat{\xi}$ to the desired precision as described in section \ref{sec:GenApproach}.

Calculating $\vec{\alpha}$ and $\frac{\partial\vec{\alpha}}{\partial\epsilon_{\alpha}}$ also appears to be computationally intensive. However, if we assume that our initial choice of parameters is not too far off the actual parameters it is justified to simply use the corresponding power spectrum for the construction of the new basis functions and keep the basis the same throughout the iterative procedure. Most of the work needed to calculate $\vec{\alpha}$ and $\frac{\partial\vec{\alpha}}{\partial\epsilon_{\alpha}}$ in each step can then be done as a one-off calculation at the start of the procedure. This is because we can write
\begin{equation}
(\vec{\alpha})_{(i,n)}=(\mat{{\cal{M}}})_{(i,n)l}C_l
\end{equation}
with
\begin{equation}
(\mat{{\cal{M}}})_{(i,n)l}=Q_{(i,n)}(\ind_1,\ind_2)\sum_m(\mat{P})_{\ind_1,lm}(\mat{P})_{lm,\ind_2}
\end{equation}
which is independent of $\epsilon$ given we keep the basis function the same. Calculating $\mat{{\cal{M}}}$ can be cumbersome but only needs to be done once. The matrix $\mat{{\cal{M}}}$ plays the same role as $\mat{M}$ relating the true power spectrum coefficients $C_l$ to the expectation values of the pseudo power spectrum estimators $\tilde{C}_l$ within the framework of PCL estimation. In fact
\begin{equation}
(\mat{M})_{l_1l_2}=(\mat{{\cal{M}}})_{(0,l_1)l_2}
\end{equation}
We can either calculate this using the analytic formula given in section \ref{subsec:PCL} or evaluate the expression directly. The analytic formula is an approximation in practice because we have to cut off the sums so if we want to calculate the matrix exactly everywhere we have to evaluate the expression directly.
We have
\begin{eqnarray}\nonumber
(\mat{{\cal{M}}})_{(0,l_1)l_2}&=&Q_{(0,l_1)}(\ind_3,\ind_4)\sum_{m_2} P_{\ind_3,l_2m_2} P_{l_2m_2,\ind_4}\\
&=&\frac{1}{2\,l_1+1}\sum_{m_1,m_2}\left(P_{l_1m_1,l_2m_2}\right)^2
\end{eqnarray}
For the other part of $\mat{{\cal{M}}}$ we obtain
\begin{eqnarray}\nonumber
(\mat{{\cal{M}}})_{(1,l_1)l_2}&=&Q_{(1,l_1)}(\ind_3,\ind_4)\sum_{m_2} P_{\ind_3,l_2m_2} P_{l_2m_2,\ind_4}\\
&=&\sum_{m_1,m_2}\left(P_{l_1m_1,l_3m_3}\frac{1}{C_{l_3}}P_{l_3m_3,l_2m_2}\right)^2
\end{eqnarray}
Similarly to the way we compute the $\beta_n$ we can implement these expressions as a succession of map generation and analysis steps using HEALPix. For example to calculate $(\mat{{\cal{M}}})_{(1,l)1}$ we generate a map consisting only of $Y^{R}_{11}$. We then mask it, analyse it, weight the resulting multipoles with $C^{-1}_{l}$, generate a new map from these multipoles, mask it again and analyse it again. The resulting multipole coefficients are squared and for each $l$ the sum over $m$ is taken. This process needs to be repeated for $Y^{R}_{10}$ and $Y^{R}_{1-1}$ to calculate the remaining contributions to $(\mat{{\cal{M}}})_{(1,l)1}$. Calculation of $(\mat{{\cal{M}}})_{(0,l_1)l_2}$ proceeds in the same way but only the first generation, masking and analysis step is required. While this can consume a considerable amount of computation time it is feasible on a state-of-the-art supercomputer. 

Making use of the ideas above the calculation of $\vec{\alpha}$ and $\frac{\partial\vec{\alpha}}{\partial\epsilon_{\alpha}}$ for each iteration step reduces to the computation of $C_l(\epsilon_{\alpha})$ and $\frac{\partial C_l}{\partial\epsilon_{\alpha}}$ which is much simpler.

Finally one can avoid the many MC runs involved in the computation of $\mat{\xi}$ in each step if the fiducial parameters are already close to the true parameters. If this is the case one can adopt the point of view that the updates change $\mat{\xi}$ only little throughout the iterative procedure and hence it is justified to simply stick to the fiducial $\mat{\xi}$ without altering the final parameter estimates considerably.

Taken together these measures make the above procedure very efficient except for a one-off calculation at the start of the iterative process.

\section{\label{app:Pixel}Effect of the finite pixel space resolution}
\begin{figure*}[htb]
\centering
\includegraphics[scale=0.25]{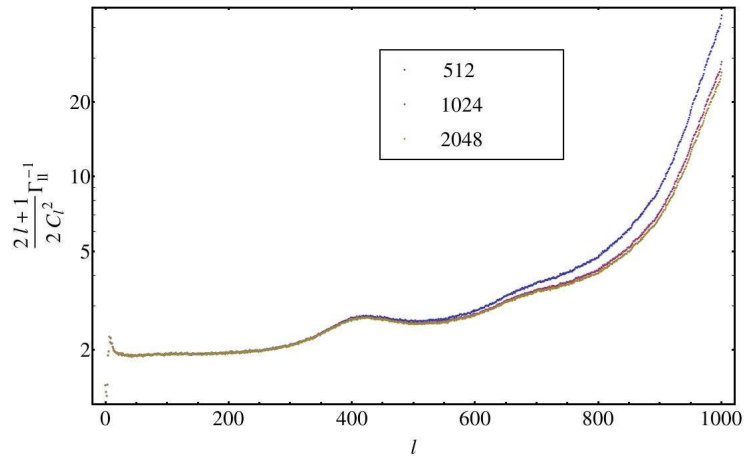}
\caption{Variance of the PCL estimator for different values of the HEALPix resolution parameter $N_{\mathrm{side}}=512,1024,2048$.}
\label{fig:errorplot}
\end{figure*}
We generally assume that the finite pixelisation of the sphere with a HEALPix grid does not have an effect on the estimates as long as one chooses a sufficiently high resolution. If we analyse a HEALPix map and determine its harmonic coefficients we have to keep in mind that the finite pixelisation can have an effect on the results. The numerical results were derived using a HEALPix grid with $N_{\mathrm{side}}=512$ analysing the maps up to $l_{\mathrm{max}}=1000$. This should provide good accuracy. However when studying high $l$ effects like the increase in the variance of the estimator when the suppression of power due to the beam is taken into account it is important to make sure that these effects are not simply due to the fact that we are not in the infinite resolution limit.
Figure \ref{fig:errorplot} shows the variance of the PCL estimator for the WMAP power spectrum suppressed by the beam as calculated from 100k MC runs for $N_{\mathrm{side}}=512,1024,2048$. We see that while there are minor changes in the variance at high $l$ going from $N_{\mathrm{side}}=512$ to $N_{\mathrm{side}}=1024$ the qualitative behaviour is the same. Increasing $N_{\mathrm{side}}$ to 2048 hardly makes a difference which suggests that this is already clearly in the infinite resolution limit for an analysis up to $l_{\mathrm{max}}=1000$.   We conclude that for high precision estimation it would be prudent to use pixelisations with $N_{\rm side} = l_{\rm max}$. 

\onecolumngrid
\bibliography{p1.bib}

\end{document}